%% file: BinaryWeave.tex
\documentclass[aps,prd,twocolumn,amsmath,amssymb,amsfonts,nofootinbib,superscriptaddress,altaffilletter]{revtex4-2}
\usepackage{graphicx}
\usepackage[unicode]{hyperref}
\hypersetup{
  bookmarksopen=true
}
\usepackage{amssymb}
\usepackage{amsmath}
\usepackage{mathtools}
\usepackage{color}
\usepackage{supertabular}
\usepackage{dcolumn}
\usepackage{multirow}
\usepackage{upgreek}
\usepackage{xcolor}
\usepackage{booktabs}  

\newif\ifshowfigs
\showfigstrue

\input{defs.tex}

\graphicspath{ {./figs/} {./} }

\usepackage{soul}

\newcommand{\useMSU}[1]{}

\begin{document}

\title[BinaryWeave: A new Sco X-1 CW-search pipeline]{Implementation of a new weave-based search pipeline for continuous gravitational waves from known binary systems}

\author{Arunava Mukherjee}
\email{arunava.mukherjee@ligo.org, arunava.mukherjee@saha.ac.in}
\affiliation{Max-Planck-Institut f{\"u}r Gravitationsphysik, Callinstr. 38, D-30167 Hannover, Germany}
\affiliation{Leibniz Universit\"at Hannover, Callinstr. 38, D-30167 Hannover, Germany}
\affiliation{Saha Institute of Nuclear Physics, 1/AF Bidhannagar, Kolkata-700064, India}

\author{Reinhard Prix}
\email{reinhard.prix@aei.mpg.de}
\affiliation{Max-Planck-Institut f{\"u}r Gravitationsphysik, Callinstr. 38, D-30167 Hannover, Germany}
\affiliation{Leibniz Universit\"at Hannover, Callinstr. 38, D-30167 Hannover, Germany}

\author{Karl Wette}
\affiliation{Centre for Gravitational Astrophysics, Australian National University, Canberra ACT 2601, Australia}
\affiliation{ARC Centre of Excellence for Gravitational Wave Discovery (OzGrav), Hawthorn VIC 3122, Australia}

\date[\relax]{compiled \today}

\begin{abstract}
  Scorpius X-1 (\Sco) has long been considered one of the most promising targets for detecting continuous
  gravitational waves with ground-based detectors.
  Observational searches for \Sco{} have achieved substantial sensitivity improvements in recent years, to
  the point of starting to rule out emission at the torque-balance limit in the low-frequency range
  $\sim\SIrange{40}{180}{Hz}$.
  In order to further enhance the detection probability, however, there is still much ground to cover for the
  full range of plausible signal frequencies
  $\sim\SIrange{20}{1500}{Hz}$, as well as a wider range of uncertainties
  in binary orbital parameters.
  Motivated by this challenge, we have developed \BinaryWeave{}, a new search pipeline for continuous
  waves from a neutron star in a known binary system such as \Sco.
  This pipeline employs a semi-coherent StackSlide $\F$-statistic using efficient lattice-based metric
  template banks, which can cover wide ranges in frequency and unknown orbital parameters.
  We present a detailed timing model and extensive injection-and-recovery simulations that illustrate that the
  pipeline can achieve high detection sensitivities over a significant portion of the parameter space when
  assuming sufficiently large (but realistic) computing budgets.
  Our studies further underline the need for stricter constraints on the \Sco{} orbital
  parameters from electromagnetic observations,
  in order to be able to push sensitivity below the torque-balance limit over the entire range of
  possible source parameters.
\end{abstract}

%
%
\maketitle

\section{Introduction}
\label{sec:introduction}

Since the first direct detection of gravitational waves from the coalescence of two stellar-mass black
holes~\cite{Abbott:2016blz}, we have observed more than 90 further gravitational-wave
events~\cite{gwtc2_1:lvc,gwtc3:lvc}.
So far, all the observed signals originated from the coalescence of binary black-hole systems, binary
neutron-star systems, and neutron-star-black-hole systems, each resulting in a short transient signal in the
gravitational-wave detectors.

A different class of gravitational-wave signals, \emph{continuous gravitational waves} (CWs) that are nearly
monochromatic and long-lasting, is yet to be observed.
Rapidly spinning neutron stars with some deviation from perfect axisymmetry are promising sources of such CWs
in the current generation of ground-based detectors, namely Advanced-LIGO (aLIGO), Advanced-VIRGO, and
KAGRA~\cite{Keith:2017mrla}.

Different physical processes within a neutron star can produce CWs, resulting in different
characteristics of the emitted signal.
For example, a non-axisymmetric deformation (or ``mountain'') on a spinning neutron star emits CWs at twice
the spin-frequency, $f=2 f\rot$, while a freely precessing neutron star will additionally emit at $f\sim
f\rot$~\cite{Keith:2017mrla}.
Oscillation modes of the internal fluids in a neutron star can also produce CWs, for example, inertial r-mode
oscillations emit at approximately $f\sim \frac{4}{3}
f\rot$ through the Chandrasekhar-Friedman-Schutz
instability~\cite{OwenEtAl:Rmode_prd1998,Nils:Rmode_apj1998,Keith:2017mrla,RajbhandariOwenEtAl:PRD2021}.

Accreting neutron stars in galactic low-mass X-ray binary (LMXB) systems are potentially strong emitters of
CWs~\cite{Wagoner:1984apj,Bildsten:1998ApJ,UshomirskyCutlerBildsten:mnras2000,cw_amxps_O3:arXiv2109.09255}, as
the accreting matter from the companion, channeled by the magnetic field of the neutron star, can result in a
substantial degree of quadrupolar non-axisymmetry of the spinning neutron
star~\cite{MelatosPayne:ApJ2005,NathanOwen:prd2013,GittinsAndersson:mnras2021}.

The accreting matter also exerts a spin-up torque on the neutron star, increasing its spin frequency $f\rot$.
Interestingly, however, the observed distribution of neutron star spin frequencies shows a pronounced cut-off
above $f\rot \sim \SI{700}{Hz}$, well below the theoretical breakup limit of realistic neutron-star equations
of state~\cite{Deepto:2008AIPC,DeeptoEtAl:Nature2003}.
Gravitational-wave emission is one of the conjectured braking mechanisms that could explain this
surprising high-frequency cut-off in the spin distribution.
According to the \emph{torque-balance} scenario, the spin-down torque due to the emission of CWs
would eventually counterbalance the accretion-induced spin-up torque.
Thus, the larger the mass accretion rate, the stronger the expected gravitational-wave emission.

\Sco{} is the brightest LMXB with one of the highest mass-accretion rates among the systems harboring a
neutron star~\cite{HasingerKlis:1989AA}. Moreover, it is relatively close to Earth, with a distance of only
$\sim$~\SI{2.8}{kpc}~\cite{ScoX1OrbPar:BradshawEtal1999}, making it one of the most promising sources of
detectable CWs~\cite{DhurandharVecchio:2001prd,ScoX1_MDC1_ChrisEtal:2015prd}.

Searching for CW signals in data from ground-based detectors is an active area of ongoing effort~\citep[e.g.,
see][for recent overviews]{Keith:2017mrla,CW-review-Keith-2022}.
We typically classify these searches (in order of decreasing computational cost) into three main categories:
\emph{all-sky} searches for unknown sources over a wide range of source parameters; \emph{directed} searches
for sources with known sky-locations and some unknown intrinsic parameters; and \emph{targeted} searches for
known pulsars, where the phase evolution of the system is assumed to be known.
Searches for \Sco{} fall into the \emph{directed} category, with a known sky position and unknown frequency,
and substantial uncertainties on some of the binary orbital parameters.

\Sco{} has long been considered one of the high-priority targets for CW searches, starting
with~\cite{ChrisAlberto:2004prd}, with further searches on initial LIGO
data~\cite{ScoX1_initialLIGO:2015prd,WhelanEtAl:prd2015,GrantEtAl_ScoX1_S6LVC:2017prd}, and more recently on
data from the Advanced LIGO
detectors~\cite{ScoX1Veterbi_O1LVC:prd2017,ScoX1CrossCorr_O1LVC:sep2017,ScoX1O2LVC:prd2019,ScoX1O2_AEI:apjl2021}.
Each successive search has improved constraints on the maximal strength of a putative CW signal from \Sco{},
with the latest constraints for the first time beating the above-mentioned \emph{torque-balance} limit in a
range of low spin frequencies
$f\rot\sim\SIrange{20}{90}{Hz}$~\cite{ScoX1O2_AEI:apjl2021}.

The past decade has seen the development and deployment of several pipelines for \Sco{} searches~\citep[e.g.,
see][for an overview]{ScoX1_MDC1_ChrisEtal:2015prd}.
Different pipelines tend to achieve different sensitivity per computing cost and degrees of
robustness against the model assumptions, such as the effect of stochastic accretion torque on the
spin-frequency evolution, i.e., the so-called \emph{spin wandering}~\cite{ArunavaChrisKeith:2018PRD}.
Recent advances in search techniques include the adaptation of the Viterbi ``hidden Markov Model'' methods to
searches for \Sco~\cite{SuvorovaEtAl_Viterbi1:2016prd,MelatosEtAl_Viterbi3:2021prd}, and the
cross-correlation \emph{CrossCorr} pipeline~\cite{DhurandharEtAl_CrossCorr:2008prd,WhelanEtAl:prd2015},
which was recently improved by using \emph{resampling} techniques~\cite{GrantEtAl_ResampCrossCorr:2018prd} as
well as efficient lattice-based template banks~\cite{WagnerEtAl:2022cqg}.

One of the open challenges for finding CWs from \Sco{} stems from the fact that most of the observed
neutron-star rotation rates in accreting LMXBs fall above $f\rot\gtrsim
\SI{300}{Hz}$~\cite{Deepto:2008AIPC,DeeptoEtAl:Nature2003}.
Given that the mass accretion rate of \Sco{} is one of the highest observed among all LMXBs, the neutron star
would have experienced a large amount of accretion-induced spin-up torque and therefore have a high spin
frequency $f\rot$.
Unfortunately, the computational cost of such a CW search grows as a substantial power of frequency
$\propto f^{3-6}$, depending on the assumed parameter space~\cite{BinaryWeave:method}.
Therefore, reaching or surpassing the torque-balance limit at higher spin frequencies becomes increasingly
challenging.

Here we present \BinaryWeave{}, a new directed \Sco{} search pipeline that employs a semi-coherent
$\F$-statistic \emph{StackSlide} approach, as outlined and analyzed in~\citet{BinaryWeave:method}.
This is achieved by extending the \Weave{} framework~\cite{IsolatedWeave}, originally developed as an
all-sky search for isolated neutron stars~\cite{2019arXiv190108998W}.
Using this framework enables us to use the fastest-available (resampling) $\F$-statistic algorithms and
efficient lattice-based metric template banks for covering the parameter space and summing $\F$-statistics
across segments.
The tuneable segment lengths and template-bank mismatch parameters allow this pipeline to translate increases
in computing budget (e.g., by using \texttt{Einstein@Home}~\cite{EinsteinAtHome} or a large computing
cluster) into improved sensitivity \cite{PrixShaltev2011:optimalStackSlide}.

\BinaryWeave{} pipeline constructs a bank of large number of templates originating from different values
of the intrinsic source parameters, e.g., spin frequency of the neutron star and orbital parameters of the
binary system. Construction of a reliable and efficient template bank maximizes the detection of a weak
signal above a predetermined threshold value. Often an increase in the number of templates increases the
detection probability but, it also requires a higher amount of computational resources. Thus, our primary
goal is to maximize the detection probability within the (varying) limitation in computing budgets. As
discussed in detail in this paper (see~\ref{characterization} and ~\ref{depth_fixed_cost}), the
construction of a template bank is key to this idea. 

There are two widely adopted general methodologies to construct template banks for GW searches, stochastic
template banks, and geometric template banks. A geometric template bank uses algorithms to place individual
templates geometrically to tile the target parameter space targeted for the search. The distance of any
two adjacent templates in multi-dimensional parameter space is dictated by the maximum amount of affordable
loss in signal-to-noise ratio which in turn is determined by the metric of the parameter space locally. 
Thus the knowledge of the parameter space metric is crucial in order to construct any geometric template
bank for any search pipeline.

The main difficulty stemmed from the fact that the \Sco{} metric changes over the parameter
space~\cite{IsolatedWeave}, while the lattice-tiling \Weave{} template-bank construction requires a
strictly constant metric.
We have solved this problem by developing a local approximation to the binary-orbital coordinates resulting in
an ``effective'' constant parameter-space metric allowing for efficient lattice tiling while satisfying good
coverage and mismatch properties.

We present and characterize the sensitivity and computational performance of \BinaryWeave{}, which
essentially realizes the predicted sensitivities in \citet{BinaryWeave:method}.
We discuss its applications for different astrophysical \Sco{} scenarios, observation setups, and computing
budgets.
We discuss some aspects of electromagnetic observations that would help to substantially alleviate the
computational challenges and improve the chances for a \Sco{} CW detection.

This paper is organized as follows: in Sec.~\ref{sec:background} we introduce the signal waveform parameters,
detection statistic and template-bank construction. Section \ref{implementation} describes the specifics of
the implementation in \BinaryWeave{}.
In Sec.~\ref{characterization} we present a detailed characterization of this new pipeline in terms of
template-bank safety as well as computing-resource requirements. Section \ref{sec:sensitivity_estimate}
presents the achievable sensitivities of this pipeline for different computational budgets, followed by
summary and outlook in Sec.~\ref{sec:summary-outlook}.

\section{Background}
\label{sec:background}

In this section we briefly introduce the concepts and notation required to understand the context of this
paper (closely following \cite{BinaryWeave:method}), namely the CW signal waveform and its parameters, the
detection statistics used and the basics of metric template-bank construction.

\subsection{Signal waveform and parameters}
\label{subsec:cw_param}

The time-dependent strain of a CW signal exerted on a gravitational-wave detector is $h(t; \A,\lambda)$,
where $t$ is the arrival time of a wavefront at the detector.
The set of four \emph{amplitude parameters} $\A$ consists of the overall amplitude $h_{0}$, the
inclination angle $\cos\iota$, polarization angle $\psi$, and the initial phase $\phi_{0}$.
The \emph{phase-evolution parameters} $\lambda$ determine the waveform phase $\phi(t;\lambda)$ as a
function of time at a given detector.
The phase evolution at the detector is determined by the source-frame frequency evolution $f(\tau)$
(dependent on the intrinsic spin-evolution of the neutron star), and by the R{\o}mer delay affecting the
arrival-time $\tau(t)$ at the detector, due to the relative motion of detector and neutron star, and
(if it is in a binary system) the star's intrinsic motion around its companion star.

The neutron-star spin typically changes slowly and can therefore be represented by a Taylor-expansion
around a \emph{reference time} $\tref$, resulting in a source-frame phase model of the form
\begin{equation}
  \label{eqn:srcframe_cw_phase}
  \phi^{\mathrm{src}}(\tau) = 2\pi \left[f(\tau - \tref) + \frac{1}{2}\dot{f}(\tau - \tref)^2 + \ldots\right],
\end{equation}
with $f$ denoting the source-frame gravitational-wave frequency at $\tref$ and its higher-order derivatives,
or ``spindown parameters'', $f^{(k)}\equiv \left.df^{k}/d\tau^{k}\right|_{\tref}$.

The waveform arrival time $\tau(t)$ from the source frame $\tau$ to the detector frame $t$ is determined by
the detector location and source sky-position (e.g., right ascension and declination), and by the orbital parameters
describing the intrinsic neutron-star motion if it is in a binary system \cite{roy_a_e_orbital_2005}.
These consist at a minimum of the orbital projected semi-major axis $\asini$, the period $\Porb$
(or equivalently the {\it mean} orbital angular velocity $\Omega \equiv {2\pi}/{\Porb}$), and a reference
time of the orbit, such as the time of ascending node $t\asc$.
These parameters would fully describe the time delay in a circular orbit, while for eccentric orbits we
additionally require the eccentricity $e$ and a rotation angle, such as the argument of periapsis $\omega$.
For systems with small eccentricity a common reparametrization uses \emph{Laplace-Lagrange} parameters
$\kappa$ and $\eta$ instead, defined as
\begin{equation}
  \kappa \equiv e\,\cos\omega,\quad
  \eta \equiv e\,\sin\omega.
\end{equation}
In the small-eccentricity limit one can relate the time of periapsis $\tp$ to the time of ascending node
$t\asc$ via
\begin{equation}
  \label{eqn:eccentricity}
  \tp - t\asc = \frac{\omega}{\Omega}.
\end{equation}
Explicit expressions for the resulting phase model can be found in
\cite{SemicohStrategyScoX1:Chris,BinaryWeave:method}.

\subsection{Detection statistics}
\label{subsec:detstat}

The gravitational-wave strain $x(t)$ observed in a detector in the presence of a signal and additive noise
$n(t)$ can be written as $x(t) = n(t) + h(t;\A,\lambda)$.
The detection problem therefore corresponds to distinguishing the pure-noise hypothesis, i.e., $h(t) = 0$,
from the signal hypothesis with non-vanishing $h(t)$.
The standard likelihood-ratio approach can be used to test different templates $h(t;\A,\lambda)$ against
the data, with a common simplification consisting in the analytic maximization over amplitude parameters,
first shown in \citet{Fstat:JKS}, resulting in the $\F$-statistic.
While this approach is not strictly optimal compared to Bayesian marginalization
\cite{ReinhardBadri_FstatVsBstat:2009cqg}, it requires far less computing cost per template $\lambda$ and is
therefore the current best choice for computationally-constrained wide parameter-space searches.

We denote the (coherent) statistic as $2\Fcoh(x;\lambda)$, which depends on the data $x$ and the
phase-evolution parameters $\lambda$ of the waveform template for which the statistic is computed.
In Gaussian noise this statistic follows a non-central $\chi^{2}$ distribution with four degrees of freedom
and a non-centrality parameter $\coh{\rho}^{2}(\A\sig, \lambda\sig; \lambda)$,
where $\A\sig$ and $\lambda\sig$ are the (unknown) signal amplitude- and phase-evolution parameters, while
$\lambda$ are the template phase-evolution parameters.
The expectation value of the coherent $\Fcoh$-statistic is
\begin{equation}
  \label{eqn:Fstat_coh}
  E\left[2\Fcoh(x;\lambda)\right] = 4 + \coh{\rho}^{2}(\A\sig, \lambda\sig; \lambda).
\end{equation}
The noncentrality parameter $\coh{\rho}^2$ characterizes the \emph{signal power} in a given template, and in
the coherent case its square-root $\coh{\rho}$ is also known as the signal-to-noise ratio (SNR) for the
coherent $\Fcoh$-statistic.

For wide parameter-space searches (such as for \Sco{}), some or all of the signal phase-evolution
parameters $\lambda\sig$ are unknown, constrained only to fall in some astrophysically-informed
parameter space $\lambda\sig \in \Dop$.
The required number of templates to cover a parameter space $\Dop$ using a coherent statistic $\Fcoh$ grows
rapidly as a function of the coherent integration time, which makes such searches effectively computationally
impossible.
Consequently, the best achievable sensitivity at a finite computational cost is typically achieved using
semi-coherent statistics, as first shown in \citet{SemiCoh_sens:Brady-etal} and analyzed in more detail in
\cite{PrixShaltev2011:optimalStackSlide}.

The \BinaryWeave{} pipeline as an extension of \Weave{} \cite{IsolatedWeave} is based on the
standard \emph{StackSlide} \cite{MendLand2005:StcHgSrSNStt} semicoherent approach using summed
$\Fcoh$-statistic over shorter coherent segments.
The total observation live-time $T\obs$ is divided into $\Nseg$ shorter segments of duration $\Tseg$, i.e., in an
ideal uninterrupted observation one would have $T\obs = \Nseg\,\Tseg$.
The semicoherent $\Fsco$-statistic is defined as the sum of the coherent per-segment $\Fcoh$-statistics over
all $\Nseg$ segments, i.e.,
\begin{equation}
  \label{eqn:coh_to_semicoh}
  2\Fsco(x;\lambda) \equiv \sum^{\Nseg}_{\ell = 1} 2\Fcoh_{\ell}(x;\lambda).
\end{equation}
This statistic follows a (non-central) $\chi^2$ distribution with $4\Nseg$ degrees
of freedom and a noncentrality parameter (or \emph{signal power}) given by
\begin{equation}
\label{eqn:rho_semicoh}
\sco{\rho}^2(x;\lambda) = \sum^{\Nseg}_{\ell = 1} \coh{\rho}^2_{\ell}(x;\lambda),
\end{equation}
so the expectation value of $2\Fsco$ is
\begin{equation}
  \label{eqn:Fstat_semicoh}
  E\left[2\Fsco(x;\lambda)\right] = 4\Nseg + \sco{\rho}^{2}(\A\sig, \lambda\sig; \lambda).
\end{equation}

\subsection{Template banks and parameter-space metrics}
\label{subsec:template_mismatch}

In order to systematically search a given parameter space $\Dop$, we need to populate it with a
finite number of templates $\lambda\tmpl \in \Dop$.
The set of all templates $\{\lambda\tmpl\}$ is referred to as the template bank, which is a discrete sampling
of $\Dop$, i.e., $\{\lambda\tmpl\} \subset \Dop$.
Due to this discretization of $\Dop$, a signal with parameters $\lambda\sig \in \Dop$ will not fall on an
exact template, resulting in a loss of recovered signal power $\rho^2$ at a template $\lambda\tmpl$, which is
quantified in terms of the \emph{mismatch} $\mis_0$, defined as the relative loss of signal power
\begin{equation}
  \label{eqn:mismatch_semicoh}
  \mis_{0}(\A\sig, \lambda\sig; \lambda\tmpl) \equiv \frac{{\rho}^{2}(\A\sig, \lambda\sig; \lambda\sig)
    - {\rho}^{2}(\A\sig, \lambda\sig; \lambda\tmpl)}{{\rho}^{2}(\A\sig, \lambda\sig; \lambda\sig)},
\end{equation}
which is a bounded function within $\mis_{0} \in [0, 1]$.

Assuming a small offset $d\lambda \equiv \lambda\tmpl - \lambda\sig$ between signal and template and neglecting the
dependence on the (unknown) signal amplitude parameters $\A\sig$ \cite{Reinhard_MultiDetFstat}, one can define
the parameter-space (phase-) metric $g_{ij}$ in terms of the truncated quadratic Taylor expansion:
\begin{equation}
  \label{eqn:metric_mismatch}
  \mis(\lambda\sig; \lambda\tmpl) = g_{ij}(\lambda\sig) \,d\lambda^{i} d\lambda^{j}\,,
\end{equation}
with implicit summation over the repeated indices $i, j = 1, \ldots, n$, where $n$ is the number of
template-bank dimensions.

The mismatch $\mis$ represents the squared \emph{distance} corresponding to the parameter offsets
$d\lambda$, and the metric $g_{ij}$ defines a distance measure on the parameter space.
As a result, one can express the \emph{bulk} number of templates $\N_{\Dop}$ in an $n$-dimensional lattice
template bank covering the parameter space $\Dop$ with maximum mismatch $\mmax$ (corresponding to the squared
covering radius of the lattice) \cite{SemiCoh_sens:Brady-etal,lattices:Reinhard} as
\begin{equation}
  \label{eqn:number_templates}
  \N_{\Dop} = \theta_n\,\mmax^{-n/2}\,\int_{\Dop} \sqrt{\det g(\lambda)}\,d^n\lambda,
\end{equation}
in terms of the lattice-specific normalized thickness $\theta_n$; a thinner lattice will cover the same volume with fewer templates.
This bulk template number ignores any extra \emph{padding} typically required to fully cover the boundary
$\partial\Dop$ of the parameter space $\Dop$, which tends to increase the total number of templates in
practice \citep[e.g., see][]{Wette_2014PRD,IsolatedWeave}.

The metric allows for a simple estimate of the approximate \emph{scale} of the template-bank resolution
along single coordinates via
\begin{equation}
  \label{eq:4}
  \delta\lambda^i = 2\sqrt{\frac{\mmax}{g_{ii}}}\,,
\end{equation}
which is obtained from Eq.~\eqref{eqn:metric_mismatch} by assuming a single nonzero offset along one
coordinate axis $\delta\lambda^i$. In a one-dimensional template-bank grid, the factor of two accounts for the
fact that the maximum mismatch $\mmax$ would be attained at the mid-point between two lattice templates.
The true higher-dimensional grid spacings will typically be larger than this estimate, however, due to
potential nonzero cross-terms $g_{ij}$ that come into play when considering generic offsets, as well as using
other lattice structures than a simple rectangular grid along coordinate axes.

A somewhat complementary grid-scale estimate can be obtained from considering the extents of the
\emph{bounding box} \cite{BinaryWeave:method} around a metric ellipse of constant mismatch
Eq.~\eqref{eqn:metric_mismatch}, namely
\begin{equation}
  \label{eq:5}
  D\lambda^i = 2\sqrt{\mmax\,(g^{-1})^{ii}}\,,
\end{equation}
where $g^{-1}$ is the inverse matrix of the metric $g$.
Contrary to Eq.~\eqref{eq:4}, this fully takes into account parameter correlations, but will generally result
in an overestimate of the actual lattice grid spacing~\cite{2013PhRvD..87h4057S, Reinhard_MultiDetFstat}.

The coherent phase metric $\coh{g}_{ij}(\lambda)$ at a parameter-space point $\lambda$ (ignoring the (unknown)
signal amplitude parameters $\A\sig$) can be shown \cite{SemiCoh_sens:Brady-etal,Reinhard_MultiDetFstat}
to be expressible directly in terms of derivatives of the signal phase $\phi(t; \lambda)$, namely
\begin{equation}
  \label{eqn:coh_phase_metric}
  \coh{g}_{ij}(\lambda) = \avg{\partial_{i}\phi(\lambda)\,\partial_{j}\phi(\lambda)}
  - \avg{\partial_{i}\phi(\lambda)}\,\avg{\partial_{j}\phi(\lambda)}\,,
\end{equation}
where $\partial_{i}\phi(\lambda) \equiv \partial\phi/\partial\lambda^{i}$ and $\avg{Q}$ denotes
time averaging of a quantity $Q$ over the coherent duration $\Delta T$, i.e.,
$\avg{Q} \equiv ({1}/{\Delta T}) \int_{t0}^{t0+\Delta T} Q(t)\,dt$.

The corresponding semicoherent metric $\sco{g}_{ij}(\lambda)$ at a point $\lambda$ can then be obtained
\cite{brady_searching_2000} as the average over segments, namely
\begin{equation}
  \label{eqn:semicoh_phase_metric}
  \sco{g}_{ij}(\lambda) = \frac{1}{\Nseg} \sum_{\ell = 1}^{\Nseg} \coh{g}_{\ell, ij}(\lambda),
\end{equation}
where $\coh{g}_{\ell, ij}$ is the coherent metric of segment $\ell$.

\subsection{\Sco{} parameter-space metric}
\label{subsec:binary_metric}

A number of rapidly spinning neutron stars in LMXB systems are found to be in (approximate) spin equilibrium
\cite{CoeEtalSpinEqui:2022mnras,HaskellPatruno:2011apjl,InsuEtalSpinEqui:1997apjl}.
According to the gravitational-wave torque-balance hypothesis, the total amount of accretion-induced spin-up
torque would be counter-balanced by the braking torques due to the emission of CWs and electromagnetic
radiation~\cite{Bildsten:1998ApJ}.
This keeps the system in \emph{approximate} torque balance, with random fluctuations in spin frequency due the
stochastic nature of the accretion flows, which is known as
\emph{spin wandering}~\cite{BildstenEtal:1997ApJS,ArunavaChrisKeith:2018PRD}.

Similar to previous studies, and following \cite{BinaryWeave:method}, we therefore assume a constant
intrinsic signal frequency $f$ with no long-term drifts, i.e., $f^{(k\ge1)}=0$, and we tackle the
spin-wandering effect by limiting the maximal segment length $\Tseg$ such that the frequency resolution is
still too coarse for any spin wandering effect to move the signal by more than one frequency
bin.

We can therefore use the following physical phase-evolution parameters describing the CW waveforms
\begin{equation}
  \label{eqn:coord_phys}
  \lambda = \{f, \asini, t\asc, \Omega, \kappa, \eta \},
\end{equation}
and assuming the small-eccentricity limit for \Sco{}, i.e., $e \ll 1$, the approximate CW phase model
\cite{SemicohStrategyScoX1:Chris,BinaryWeave:method} can be written as
\begin{equation}
  \label{eqn:binary_phasemodel}
  \frac{\phi(t;\lambda)}{2\pi} \approx f \Delta t -
  f \asini \left[\sin\Psi + \frac{\kappa}{2}\sin2\Psi - \frac{\eta}{2}\cos2\Psi \right],
\end{equation}
where $\Delta t\equiv t - \tref$ and the orbital phase $\Psi(t)$ is given by
\begin{equation}
  \label{eq:2}
  \Psi(t) = \Omega\, (t -  t\asc).
\end{equation}

From the explicit expression Eq.~\eqref{eqn:binary_phasemodel} of the phase one can obtain the phase
derivatives $\partial_i \phi$ with respect to the parameter-space coordinates $\lambda^i$, and time-averaging
yields the coherent metric components $\coh{g}_{\ell,ij}$ for each segment $\ell$ according to
Eq.~\ref{eqn:coh_phase_metric}.
The semi-coherent metric $\sco{g}_{ij}$ is then obtained by averaging over segments following
Eq.~\ref{eqn:semicoh_phase_metric}.

In the template-bank construction the metric will typically be computed numerically starting from the analytic
expressions for the phase derivatives.
However, it is important to also consider approximate analytic expressions for these metrics, in order to
better understand their properties.
As discussed in \cite{SemicohStrategyScoX1:Chris,BinaryWeave:method}, analytic approximations can be found in
the two limiting cases: \emph{short segments} where $\Tseg\ll\Porb$, or \emph{long segments} where $\Tseg\gg \Porb$.
Longer segments will result in better sensitivity but also higher computational cost.
The results in \cite{BinaryWeave:method} indicate that using a realistically large computational budget,
semi-coherent $\Fsco$-statistic searches for \Sco{} can afford segments substantially longer than
$\Porb\sim\SI{19}{\hour}$.
Therefore we will only discuss the \emph{long-segment} limit here, for which the nonzero elements in the
analytic approximation to the semi-coherent metric are found \cite{BinaryWeave:method} as
\begin{equation}
  \label{eqn:ls_semicoh_metric}
  \begin{aligned}
    \sco{g}_{ff} &= \pi^{2} \frac{\Tseg^{2}}{3} \,, \\
    \sco{g}_{\asini\asini} &= 2\pi^{2} f^{2} \,, \\
    \sco{g}_{\Omega \Omega} &= 2\pi^{2} (f \asini)^{2}
    \left(\frac{\Tseg^{2}}{12} + \av{\Delta\ma^{2}} \right) \,, \\
    \sco{g}_{t\asc t\asc} &= 2\pi^{2} (f \asini \Omega)^{2} \,, \\
    \sco{g}_{\Omega t\asc} &= \sco{g}_{t\asc \Omega} = -2\pi^{2} (f \asini)^{2}\,
    \Omega \, \av{\Delta\ma} \,, \\
    \sco{g}_{\kappa \kappa} &= \sco{g}_{\eta \eta} = \frac{\pi^{2}}{2} (f \asini)^{2}\,,
  \end{aligned}
\end{equation}
where $\Delta\mal \equiv t\midl - t\asc$ is the time offset between the midpoint $t\midl$ of segment
$\ell$ and the ascending node $t\asc$, and where $\av{Q}$ denotes averaging over segments, i.e.,
$\av{Q} \equiv (1/\Nseg)\sum^{\Nseg}_{\ell = 1} Q_{\ell}$.
Note that the coherent per-segment metric $\coh{g}_{\ell,ij}$ can simply be read-off these expressions as the
special case $\ell=\Nseg=1$.

There are two important aspects to consider about this metric:
\begin{enumerate}
\item The metric components still depend on the search parameters $f,\asini,\Omega$ and $t\asc$ and are
  therefore \emph{not constant} over the parameter space. This is an obstacle to constructing a lattice
  template-bank, which will be dealt with in Sec.~\ref{implementation}.
\item There is little \emph{refinement} of the semi-coherent metric compared to the per-segment coherent
  resolution, in fact most components do not depend on the number of segments (i.e., the total duration 
  of data used for the searches) for a fixed duration of coherent segment $\Tseg$, except for
  $\sco{g}_{\Omega\Omega}$ (via $\av{\Delta\ma^2}$) and $\sco{g}_{\Omega t\asc}$ (via $\av{\Delta\ma}$).
\end{enumerate}
In order to simplify the expression, we can make use of the gauge freedom in $t\asc$, which is only defined up
to an integer multiple of the period $\Porb$, i.e.,
\begin{equation}
  \label{eq:1}
  t\asc' = t\asc + n\,\Porb,\quad\text{for}\quad n\in\mathbb{Z},
\end{equation}
describes the same physical orbit, as seen in Eqs.~\eqref{eqn:binary_phasemodel},\eqref{eq:2}.
Given the long-segment assumption $\Tseg\gg\Porb$, the total observation time will satisfy this even more
strongly, i.e., $T\obs\ge\Nseg\,\Tseg \gg \Porb$.
One can therefore chose a gauge $t\asc\approx \av{t\mid}$ such that $\av{\Delta\ma}\approx0$, removing the
only nonzero off-diagonal component $g_{\Omega t\asc}$.
Further, assuming gapless segments one can show \cite{BinaryWeave:method} that in this case
\begin{equation}
  \label{eq:3}
  \sco{g}_{\Omega\Omega} = \frac{\pi^{2}}{6} (f \asini)^{2}\,(\Nseg\,\Tseg)^2,
\end{equation}
in other words, only the semi-coherent resolution in $\Omega$ increases with the number of segments, while all
other parameters have the same metric resolution per segment and in the semi-coherent combination.
This point will be further discussed in Sec.~\ref{implementation} on the details of the \BinaryWeave{}
implementation.

\subsection{Lattice-tiling template banks}
\label{subsec:template_and_lattice}

The template-bank construction in \BinaryWeave{} is directly inherited from \Weave{}, described in full
detail in \cite{Wette_2014PRD,IsolatedWeave}, therefore we only provide a short overview here.
The basic inputs to the lattice-tiling algorithm are the parameter-space coordinates $\{\lambda^i\}$,
boundaries defining $\Dop$, and the corresponding template-bank metric, which must be constant over the search
space.
The code can use a coordinate-transformation to \emph{internal} coordinates if the metric is expressed in
different coordinates than the standard CW waveform parameters described in Sec.~\ref{subsec:cw_param}.
Based on these inputs, together with a maximum-mismatch parameter $\mmax$ and a choice of lattice type, the
algorithm constructs a template-bank lattice with covering radius $\sqrt{\mmax}$ tiling the parameter space
(and ensuring appropriate covering of the boundaries).

There are two main \emph{modes} semi-coherent statistics can be computed over the set of segments:
\emph{interpolating} and \emph{non-interpolating}.
As mentioned in Sec.~\ref{subsec:binary_metric}, the semi-coherent template bank requires a finer
resolution in $\Omega$ to compute $\Fsco$ than the per-segment template banks to compute $\Fcoh_\ell$ at given
maximum-mismatch $\mmax$.
This can be used to save computing power, by using coarser per-segment template banks together with a
nearest-neighbor interpolation when picking per-segment $\Fcoh_\ell$ to sum in Eq.~\eqref{eqn:coh_to_semicoh}.
The details and effects of such an interpolating StackSlide approach are discussed in
\cite{PrixShaltev2011:optimalStackSlide}.
The simpler, yet generally more computationally expensive, method consist in using the same semi-coherent
fine grid over all segments, such that Eq.~\eqref{eqn:coh_to_semicoh} can directly be computed without any interpolation.

The amount of computing-cost savings due to interpolation depends on the refinement factor between coherent
and semi-coherent metrics, which in the case of the \Sco{} metric is only linear in $\Nseg$ if including
$\Omega$ in the template bank, and unity otherwise, as discussed in Sec.~\ref{subsec:binary_metric}.
The expected sensitivity gains by using interpolation in this case would therefore be modest and partially
reduced by the extra mismatch incurred due to interpolation itself \citep[see][]{PrixShaltev2011:optimalStackSlide}.
Furthermore, it is more difficult to find optimal setup parameters for an interpolating setup, given there are
two mismatch parameters $\{\mmaxcoh,\mmaxsco\}$ to tune rather than a single $\mmax$, in addition to the
number $\Nseg$ and length $\Tseg$ of the semi-coherent segments.

The \Ans{} lattice is a common choice
\cite{WettePrix_2013PRD, Wette_2014PRD, IsolatedWeave, Wette_etal_2021PRD}
as a close-to-optimal template-bank lattice to use, based on earlier arguments about optimal covering
lattices \cite{lattices:Reinhard,conway_sphere_1999}.
Recent work \cite{Bruce:2021PRD,BruceShoom:2021PRD} has clarified, however, that finding the template-bank
lattice that maximizes the expected detection probability at fixed number of templates is an instance of the
\emph{quantizer} problem \cite{conway_sphere_1999}, not the \emph{covering} problem.
This changes somewhat the choice of current ``record holder'' lattice in each dimension, and reduces the
relative advantage of \Ans{} over the hyper-cubic lattice, but even in this paradigm \Ans{} remains a
close-to-optimal lattice and therefore continues to be a practically reasonable and sound choice.

\subsection{\Sco{} parameter space}
\label{subsec:sco_params}

Optical and X-ray observations tell us that \Sco{} is an LMXB system~\cite{PrendergastBurbidge:1968apjl}.
Furthermore, X-ray spectral and timing characteristics indicate that the compact object in the \Sco{} binary
system is a neutron star~\cite{HasingerKlis:1989AA}.
Observations in optical and radio bands have constrained the three orbital parameters $\asini$, $\Porb$, and
$t\asc$ of \Sco{} to different extents
\cite{ScoX1OrbPar:BradshawEtal1999,ScoX1OrbPar:FomalontEtal2001,ScoX1OrbPar:WangEtal2018}, namely
$\asini \in [1.45, 3.25]\,\si{ls}$,
$\Porb \sim \SI{68 023.86048 +- 0.0432}{\second}$ and
$t\asc \sim \SI{974 416 624 \pm 50}{GPS\,\second}$ (as used in a recent CW search \cite{ScoX1O2LVC:prd2019}.
while the spin frequency of the neutron star still remains practically unconstrained
\cite{ScoX1CygX2Pulsation2021,GalaudageEtal:2022MNRAS} to date.
We provide a table of various \Sco{} parameter-space ranges considered in this and past studies (and
searches) in Table.~\ref{tab:search_param_spaces}, which will be discussed in more detail in
Sec.~\ref{cost_fixed_depth}.

The typical life-cycle of an LMXB along with one of the highest mass-accretion rate systems indicate that
the neutron star in Sco X-1 is likely to receive a large amount of accretion-induced spin-up torque
\cite{BildstenEtal:1997ApJS} and is plausibly spinning rapidly. Most of the accreting neutron stars in
LMXB systems are observed to be spinning in the range of $\sim\SIrange{300}{600}{Hz}$, although a few of them
have also been observed at lower spin frequencies~\cite{DeeptoEtal:2003Natur, Deepto:2008AIPC}.
We explore the implications of different assumptions about the spin and orbital parameters of \Sco{} for a
wide-parameter search in subsection~\ref{cost_fixed_depth}.

\section{Flat metric approximation}
\label{implementation}

As discussed in Sec.~\ref{subsec:binary_metric}, the long-segment binary parameter-space metric
Eq.~\eqref{eqn:ls_semicoh_metric} in physical coordinates is not constant over $\{f, \asini, \Omega, t\asc\}$,
which prohibits its direct use for lattice tiling.
This represents the main obstacle to applying the \Weave{} framework to a directed binary search.

Regarding the frequency dependence, all metric components (except $g_{ff}$) scale as $f^2$, as the signal phase
at the detector is $\phi(t) \sim 2 \pi f\,\tau(t)$ and the metric \eqref{eqn:coh_phase_metric} is quadratic in phase.
This scaling is similar to the metric over the sky position
parameters, and may be mitigated in the same way~\cite[e.g.][]{singh_results_2016}:
a full search is typically broken into smaller \emph{workunits} distributed over nodes of a cluster (or
Einstein@Home), where each workunit would analyze a relatively narrow frequency band
$\lesssim \Ord{\SI{1}{Hz}}$.
We can therefore deal with the frequency dependence by simply evaluating the metric at a fixed
frequency within each narrow range, typically at the highest frequency to guarantee the given maximum-mismatch
constraint over the search band, accepting small relative changes of the mismatch distribution over the
frequency band.

A similar argument applies to $t\asc$ in the long-segment regime: due to the
gauge freedom Eq.~\eqref{eq:1}, the maximal physical uncertainty for any system would be $\Delta t\asc<\Porb\ll\Tseg$,
and can therefore be neglected in $g_{\Omega\Omega}$, as seen in Eq.~\eqref{eq:3}, which is the only metric
term that would be affected by this.

Given the narrow astrophysical uncertainties on $\Porb$ for \Sco{} (cf.\ Sec.~\ref{subsec:sco_params}),
this approach could also be used for $\Omega$, but it would be very specific to \Sco{} and might not apply to
other directed binary searches.
Furthermore, ignoring the metric changes over the astrophysical range on $\asini$ would not work well for
\Sco{}, given the currently uncertainty spans more than a factor of two.

We observe that the metric Eq.~\eqref{eqn:ls_semicoh_metric} depends on $\asini$ and $\Omega$ only via quadratic
\emph{scaling} of some components, i.e., the metric stretches or contracts along certain directions in
parameter space.
In order to absorb this scaling, we only need to assume that the metric change is negligible \emph{on the
scale $\delta\lambda$ of a lattice cell}, so we can resort to \emph{local} rescaling via the following
``pseudo'' coordinate transformation of $\{\Omega,t\asc,\kappa,\eta\}$ into:
\begin{equation}
  \label{eqn:coord_transf}
  \begin{aligned}
    \vp &\equiv \fix{\asini}\,\Omega,\\
    d\asc &\equiv \fix{\asini}\,\fix{\Omega}\,t\asc,\\
    \kappap &\equiv \fix{\asini}\,\kappa,\\
    \etap &\equiv \fix{\asini}\,\eta,
  \end{aligned}
\end{equation}
where $\fix{\asini}$ and $\fix{\Omega}$ will be treated as constant scaling parameters in derivatives.
Substituting the new coordinates in Eq.~\eqref{eqn:binary_phasemodel} results in the (orbital)
phase model
\begin{equation}
  \label{eq:new-phase-model}
  \begin{aligned}
    \frac{\phi\orb(t;\lambda)}{2\pi} &=
    - f \asini \left(\sin\Psi + \frac{\kappap}{2\fix{\asini}}\sin2\Psi - \frac{\etap}{2\fix{\asini}}\cos2\Psi \right),\\
    \Psi(t) &= \frac{\vp}{\fix{\asini}}\left(t - \frac{d\asc}{\fix{\asini}\,\fix{\Omega}}\right),
  \end{aligned}
\end{equation}
and the following \emph{approximate} phase derivatives:
\begin{equation}
  \label{eqn:phys_phasederv_latt_coord}
  \begin{aligned}
    \partial_{\vp} \phi &= - 2\pi f (t - t\asc)\, [\cos\Psi + \kappa\cos2\Psi + \eta\sin2\Psi] \,,\\
    \partial_{d\asc} \phi &= 2\pi f \,[\cos\Psi + \kappa\cos2\Psi + \eta\sin2\Psi] \,,\\
    \partial_{\kappap} \phi &= - \pi f \sin2\Psi \,,\\
    \partial_{\etap} \phi &= \pi f \cos2\Psi.
  \end{aligned}
\end{equation}
Applying the steps of Sec.~\ref{subsec:template_mismatch} this yields the following metric components (with
$\sco{g}_{ff}$ and $\sco{g}_{\asini\asini}$ unchanged from Eq.~\eqref{eqn:ls_semicoh_metric}):
\begin{equation}
  \label{eqn:ls_semicoh_metric_latt_coord}
  \begin{aligned}
    \sco{g}_{\vp\vp} &= 2\pi^{2} f^{2}\,\left(\frac{\Delta T^{2}}{12} + \av{\Delta\ma^{2}} \right) \,,\\
    \sco{g}_{d\asc d\asc} &= 2\pi^{2} f^{2} \,,\\
    \sco{g}_{\vp d\asc} &= \sco{g}_{d\asc\vp} = -2\pi^{2} f^{2} \,\av{\Delta\ma} \,,\\
    \sco{g}_{\kappap\kappap} &= \sco{g}_{\etap\etap} = \frac{\pi^{2}}{2} f^{2}\,,
  \end{aligned}
\end{equation}
which are constant over $\asini$ and $\Omega$ and are therefore suitable for lattice tiling within
the \Weave{} framework.

We are applying the coordinate transformation Eq.~\eqref{eqn:coord_transf} globally over the
search parameter space, but ignore the local changes in $\asini,\Omega$-scaling within each lattice cell.
This should be a good approximation as long as cells are small compared to the effects of changing
$\asini,\Omega$ over their respective length scales.

We have thoroughly tested the safety and effectiveness of this metric approximation for a \Sco{} search,
which is discussed in the next section.

\section{Testing and characterization}
\label{characterization}

The semi-coherent $\Fsco$-statistic in \BinaryWeave{} is computed by the well-tested \Weave{}
framework~\cite{IsolatedWeave} using the standard \textsc{LALSuite}~\cite{LAL} $\F$-statistic implementation.
The behavior of this statistic implementation in recovering signals in noise is therefore already well
understood and tested. Therefore, the only new elements requiring careful testing and characterization are the
template-bank mismatch and the computing cost. For this reason, we have limited the mismatch characterization
studies for the signal-only cases without introducing any kind of GW detector noises.

\subsection{Test setup and assumptions}
\label{sec:test-setup-assumpt}

The metric and template bank implemented in \BinaryWeave{} can in principle handle eccentricity within the
small-eccentricity approximation $e\ll 1$ of Eq.~\eqref{eqn:binary_phasemodel}, which in
\cite{BinaryWeave:method} was seen to hold up to about $e \lesssim 0.1$.
The orbital eccentricity of \Sco{} is currently poorly constrained, but expected to be close to zero due to
Roche-lobe overflow accretion \cite{ScoX1OrbPar:WangEtal2018}.
In order to simplify this first proof-of-concept study of \BinaryWeave{}, we are assuming negligible
eccentricity here and focus on purely circular orbits.
Therefore we consider a \Sco{} search parameter space $\Dop$ that is (at most) four-dimensional (4D), with
search parameters $\{f,\asini, \Porb, t\asc\}$.

The orbital period $\Porb$ for \Sco{} is constrained to about
$\Delta\Porb\sim\SI{0.04}{\second}$, compared to a period of $\Porb\sim\SI{19}{\hour}$ (cf.\ Sec.~\ref{subsec:sco_params}).
The search resolution $\delta\Omega$ (and therefore also $\delta\Porb$) in Eq.~\eqref{eq:4} is determined by
the metric (in particular $\sco{g}_{\Omega\Omega}$ of Eq.~\eqref{eq:3}) and therefore depends on the search
setup $\{\Nseg,\Tseg,\mmax\}$, the search frequency $f$, and semi-major axis $\asini$.
In particular, the resolution increases linearly with total search duration $T\obs=\Nseg\,\Tseg$, and for
longer-duration searches (e.g., $T\obs\sim \SI{6}{months}$) will often fully resolve the parameter-space
uncertainty in period, i.e., $\delta\Porb < \Delta\Porb$.
However, for coarser search setups, or assuming future improved observational constraints, it can also be
sufficient to place a single template at the mid-point of the uncertainty range, resulting in a
three-dimensional (3D) search space $\Dop$ spanning only $\{f, \asini, t\asc\}$.
In the following we will therefore consider both possibilities of 3D and 4D template banks.

In this study we are exclusively using the \emph{non-interpolating} StackSlide \Weave{} mode, which is
simpler and easier to optimize for, while expected to yield similar sensitivity for directed binary searches,
as discussed in Sec.~\ref{subsec:template_and_lattice}.
This means that the coherent segments and final semi-coherent statistic use the same template grid and there
is only a single mismatch parameter $\mmaxsco=\mmaxcoh$.

All subsequent simulations use $\F$-statistic input data split into short Fourier-transforms (SFT)
\cite{prix_f-statistic_2010} of baseline $T\sft\le \SI{250}{\second}$, which is a safe SFT length over the
\Sco{} parameter space, e.g., see Eq.(C2) in \cite{BinaryWeave:method}.
Furthermore, all simulations assume data from two detectors, namely LIGO Hanford (H1) and LIGO Livingston
(L1).

\subsection{Template-bank mismatch}

In order to ensure the validity of the constructed lattice template banks using the approximately-flat
metric constructed in Sec.~\ref{implementation}, we perform injection-recovery Monte-Carlo tests.
These tests are typically performed \emph{without noise}, i.e., searching a data stream only
containing the injected signal waveform. This allows one to directly measure signal power without noise bias
and to accurately calculate the mismatch, which is the main purpose of template bank tests.
The signal parameters for the injections are drawn uniformly from the (wider) testing \Sco{} parameter-space
$\Dop_0$ specified in Table.~\ref{tab:search_param_spaces}, with randomly drawn amplitude parameters $\A$, and
a search grid is constructed around the injection point (randomly shifted to avoid systematic alignment effects).

A good template bank should satisfy the \emph{maximal mismatch criterion} \citep[e.g.][]{lattices:Reinhard}:
the measured mismatch $\mis_0(\lambda\sig;\lambda\tmpl)$ of Eq.~\eqref{eqn:mismatch_semicoh} for any injected signal
$\lambda\sig\in\Dop$ at its ``closest'' (i.e., highest signal power $\rho^2$) template $\lambda\tmpl$ should
be less than the maximum mismatch $\mmax$ the template bank was constructed for, which can formally be written as
\begin{equation}
  \label{eqn:mismatch_criteria}
  \max_{\lambda\sig\in\Dop}\min_{\lambda\tmpl}\,\mis_0(\lambda\sig;\lambda\tmpl) \le \mmax,
\end{equation}
where the ``minimax'' formulation (constraining the nearest-template mismatch at the worst-case signal
location) implies that the mismatch is constrained for all possible signal locations.

Furthermore, an \emph{efficient} template bank should ideally place only a single template within $\mmax$ of any
signal, to avoid (computationally wasteful) over-resolution and producing excessive candidates per signal that
would require some form of clustering or follow-up (see also \cite{2019arXiv190108998W}).

We have performed a number of signal injection-recovery tests of the \BinaryWeave{} template banks for various
different search setups $\{\Nseg,\Tseg,\mmax\}$.
Here we only present a few representative examples in order to illustrate the main features of these template
banks: in Sec.~\ref{lattice_1D} we illustrate the template grids for single-parameter (1D) and
two-parameters (2D) searches, and in Sec.~\ref{sec:testing-3d-4d} we provide examples of the mismatch
distribution for 3D searches (for a non-resolved period uncertainty $\Delta\Porb$) and full 4D searches.

\subsubsection{Testing 1D and 2D lattice tilings}
\label{lattice_1D}

\begin{figure*}[htbp]
  \raggedright (a)\hspace*{\columnwidth}(b)\\[-0.5cm]
  \includegraphics[clip,width=\columnwidth]{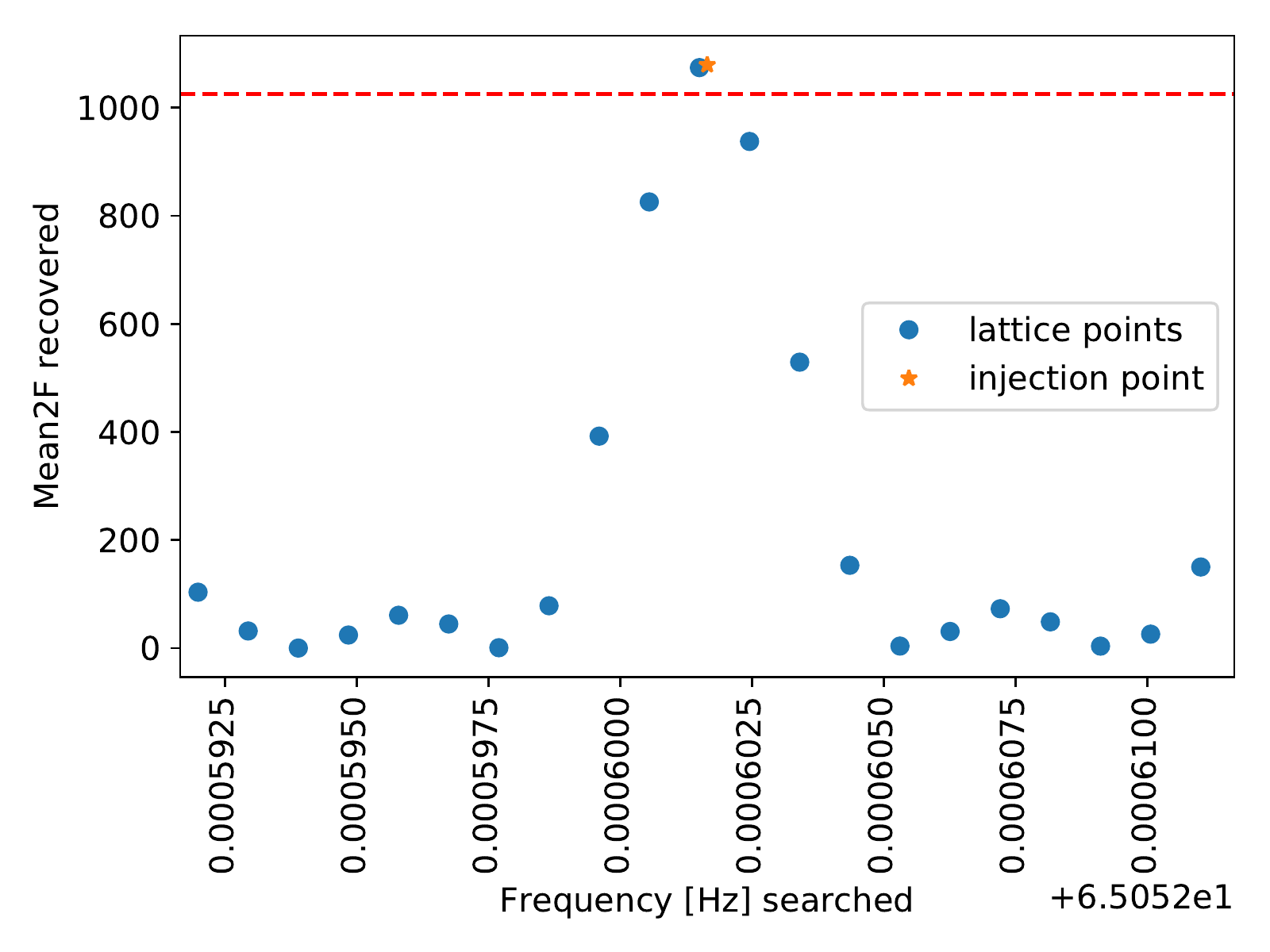}
  \includegraphics[clip,width=\columnwidth]{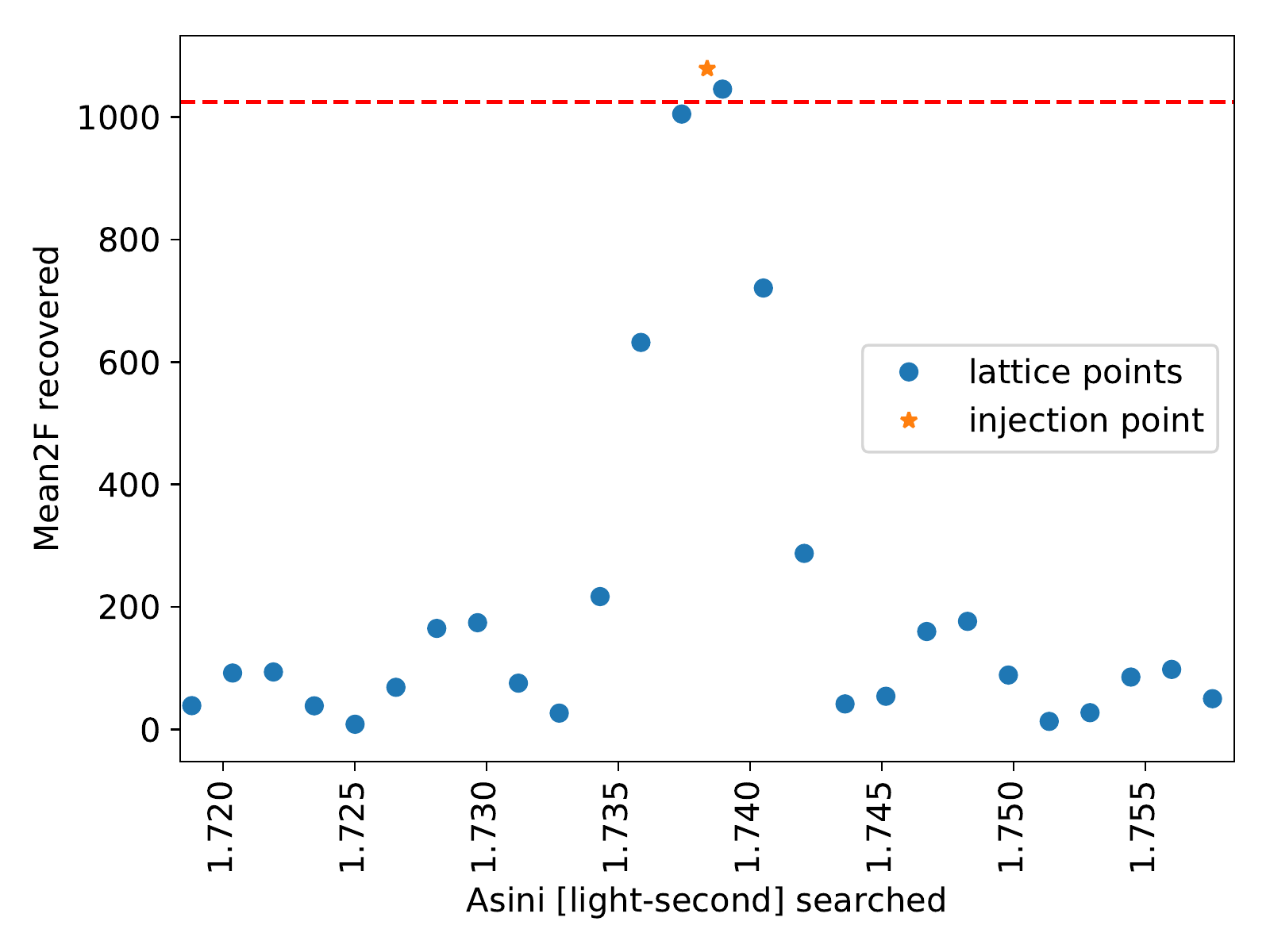}\\[0.3cm]

  \raggedright (c)\hspace*{\columnwidth}(d)\\[-0.5cm]
  \includegraphics[clip,width=\columnwidth]{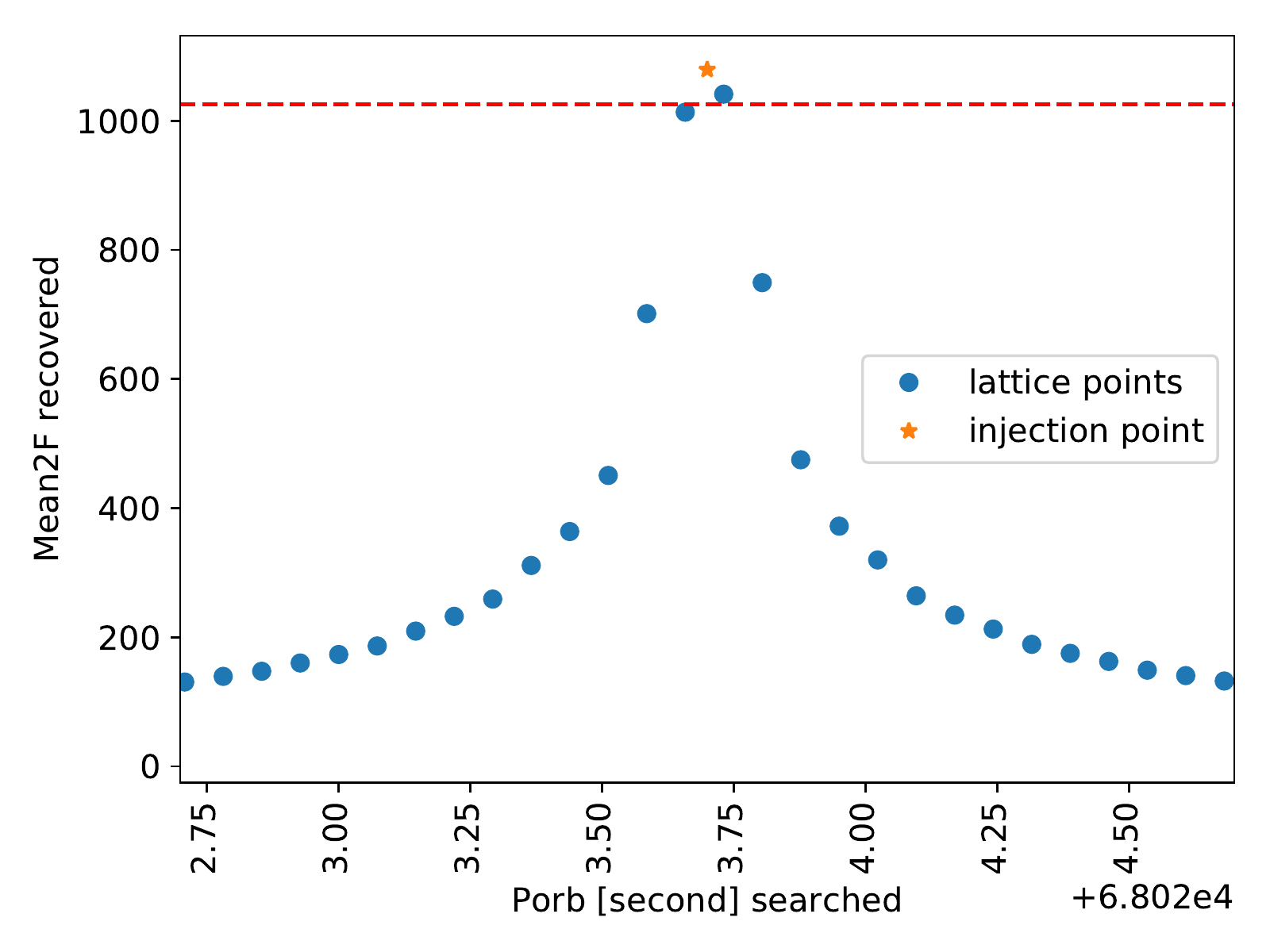}
  \includegraphics[clip,width=\columnwidth]{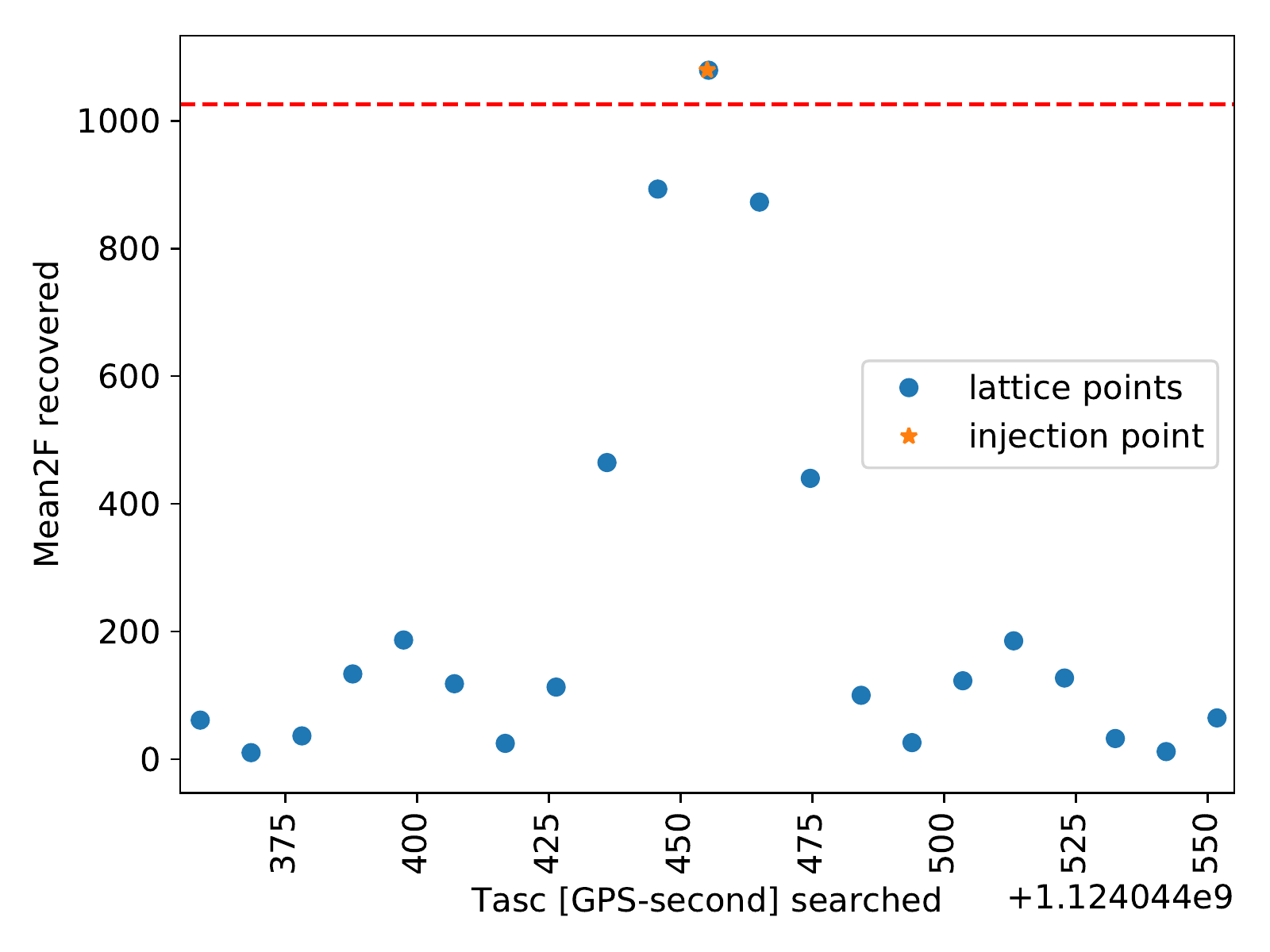}\\[0.3cm]

  \caption{Illustration of 1D template-bank searches around a noiseless signal injection, with the respective
    three remaining search parameters fixed to the injected signal.
    The filled circles mark the placement of templates and their corresponding measured $\Fsco$-statistic
    values, while the star marks the signal injection point with its corresponding perfect-match
    $\Fsco$-statistic.
    The template bank was constructed for a maximum mismatch of $\mmax = 0.05$, with $\Nseg=120$ segments of
    $\Tseg=\SI{3}{\day}$.
    The horizontal dashed line denotes the $\Fsco$-value corresponding to the maximum-mismatch criterion
    Eq.~\eqref{eqn:mismatch_criteria} relative to the injected signal power.
  }
  \label{fig:templates_1param}
\end{figure*}

In order to illustrate and visualize the lattice tiling, we first consider simple one- and two-dimensional
lattice cases, which also serve as a basic sanity check for the template bank construction.
The 1D searches are performed along all four coordinate axis in a neighborhood around the signal
injection, with the three remaining parameters fixed to the injection values, with one example shown in
Fig.~\ref{fig:templates_1param}.
\begin{figure*}[htbp]
  \includegraphics[clip,width=\columnwidth]{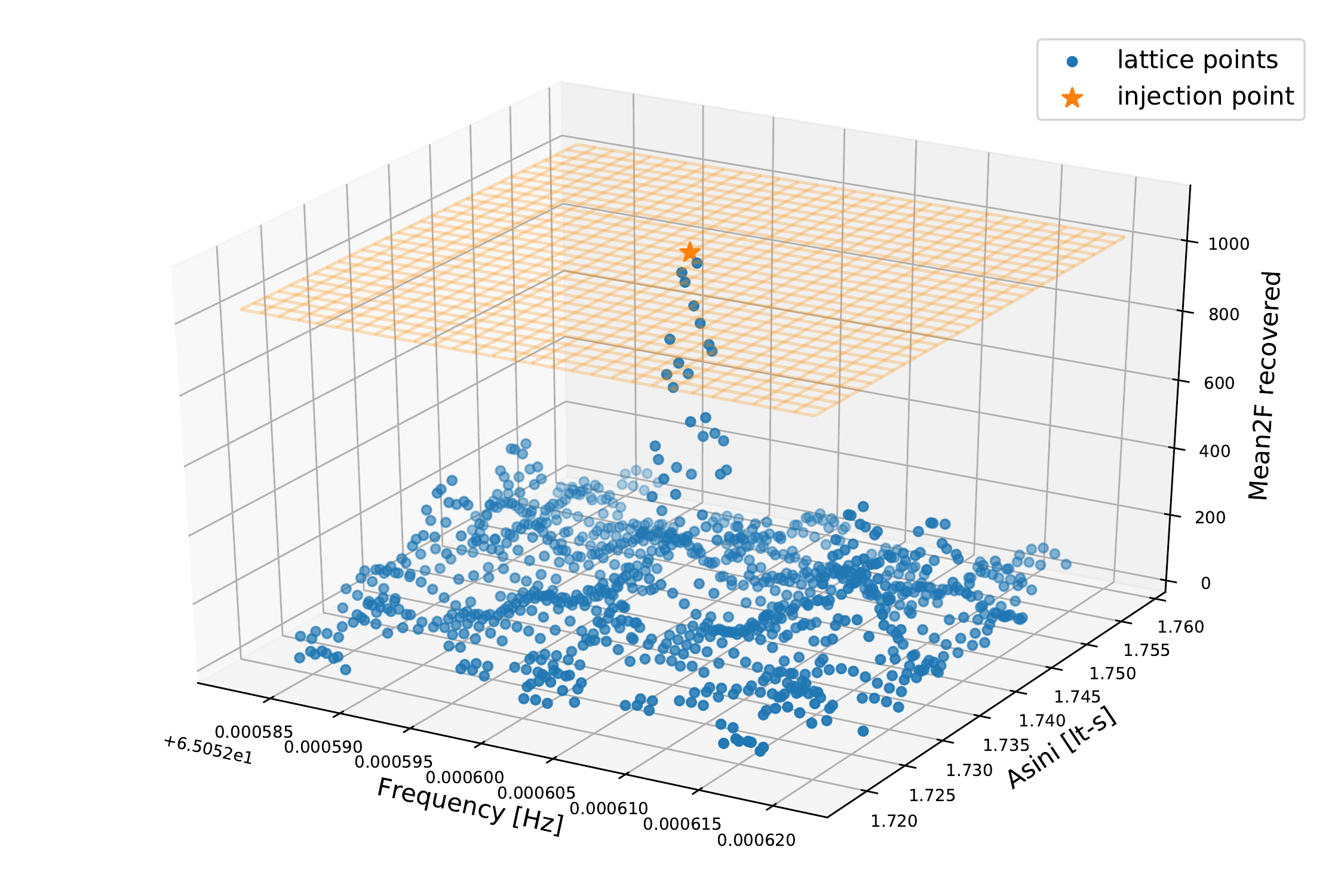}
  \includegraphics[clip,width=\columnwidth]{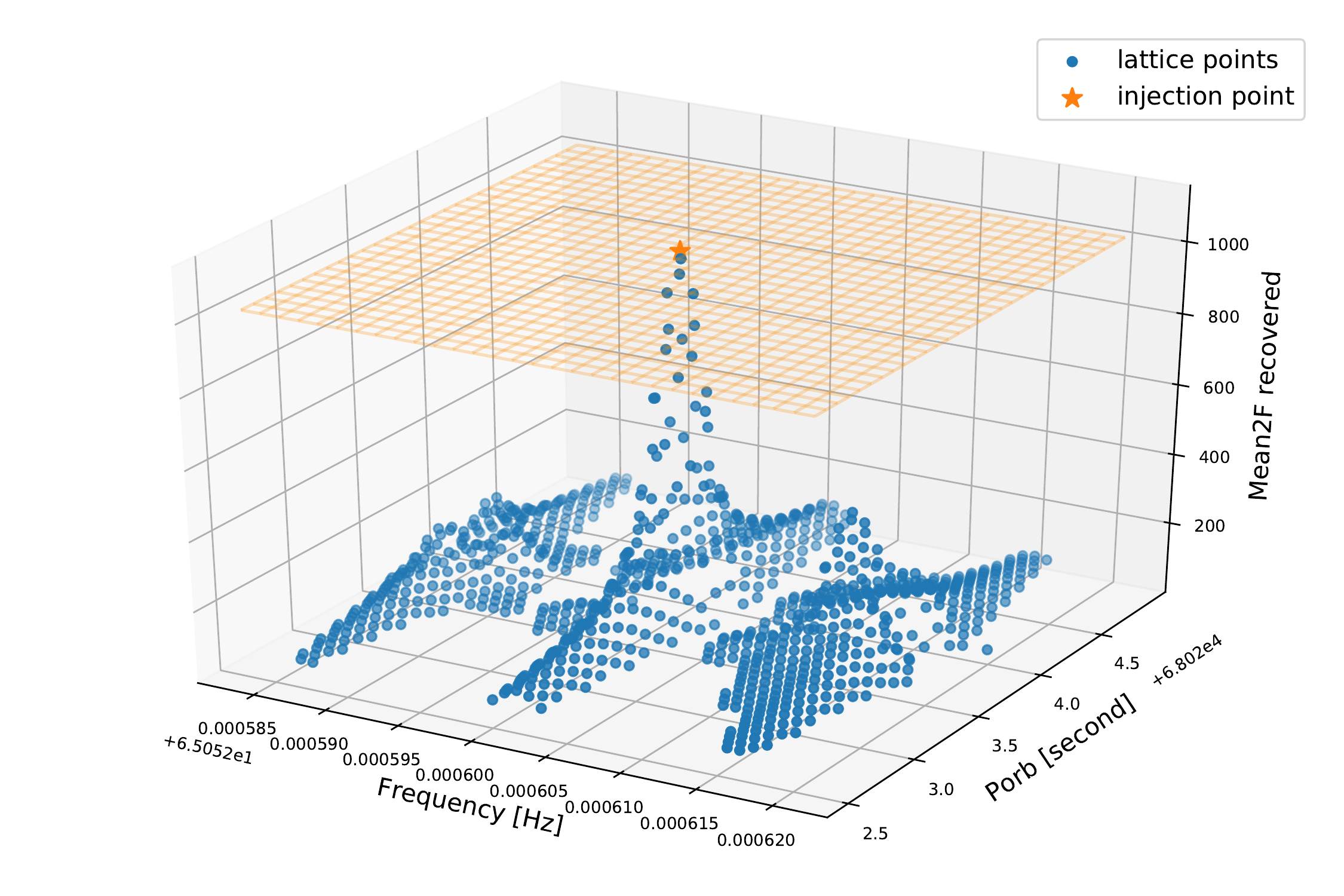}\\
  \includegraphics[clip,width=\columnwidth]{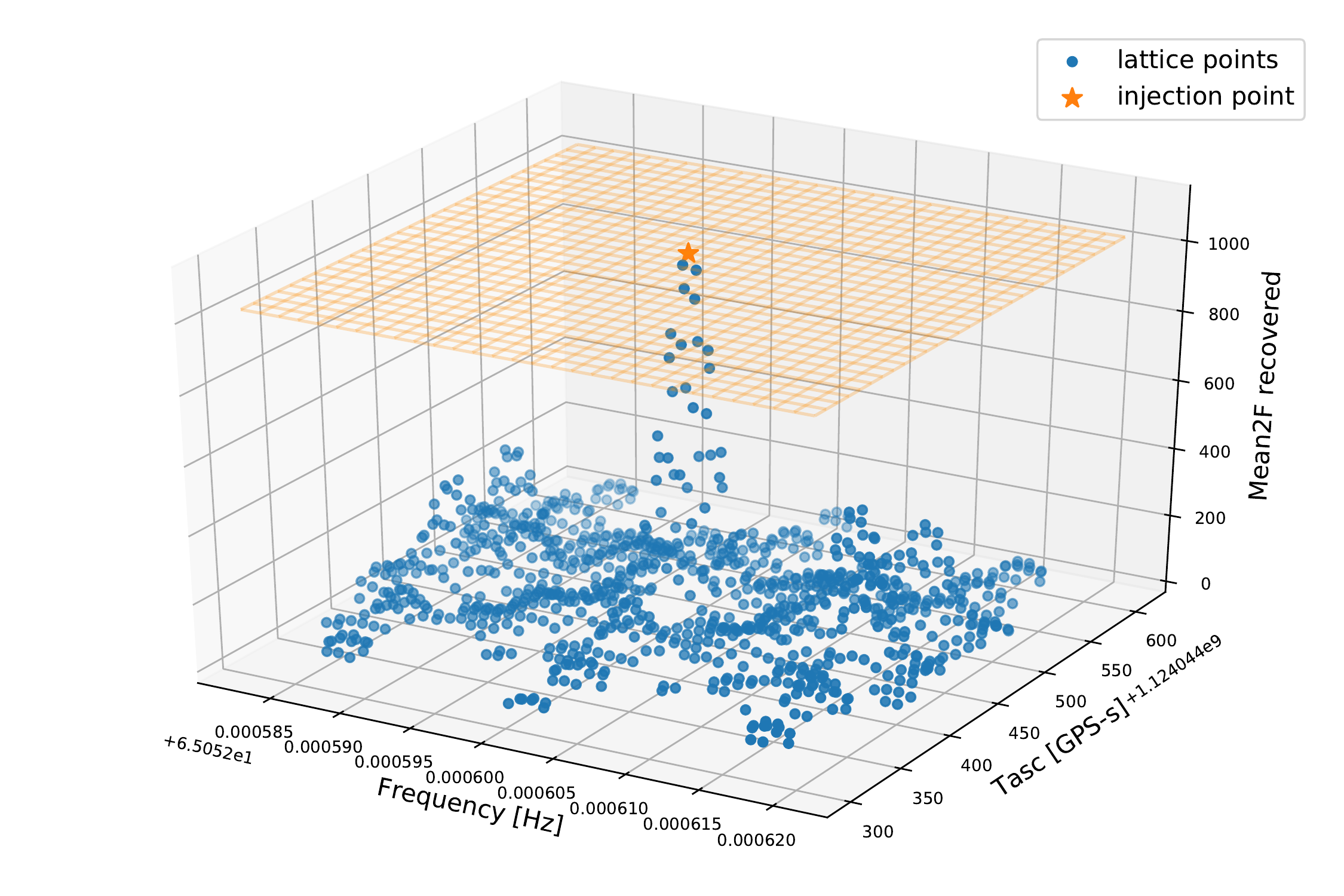}
  \includegraphics[clip,width=\columnwidth]{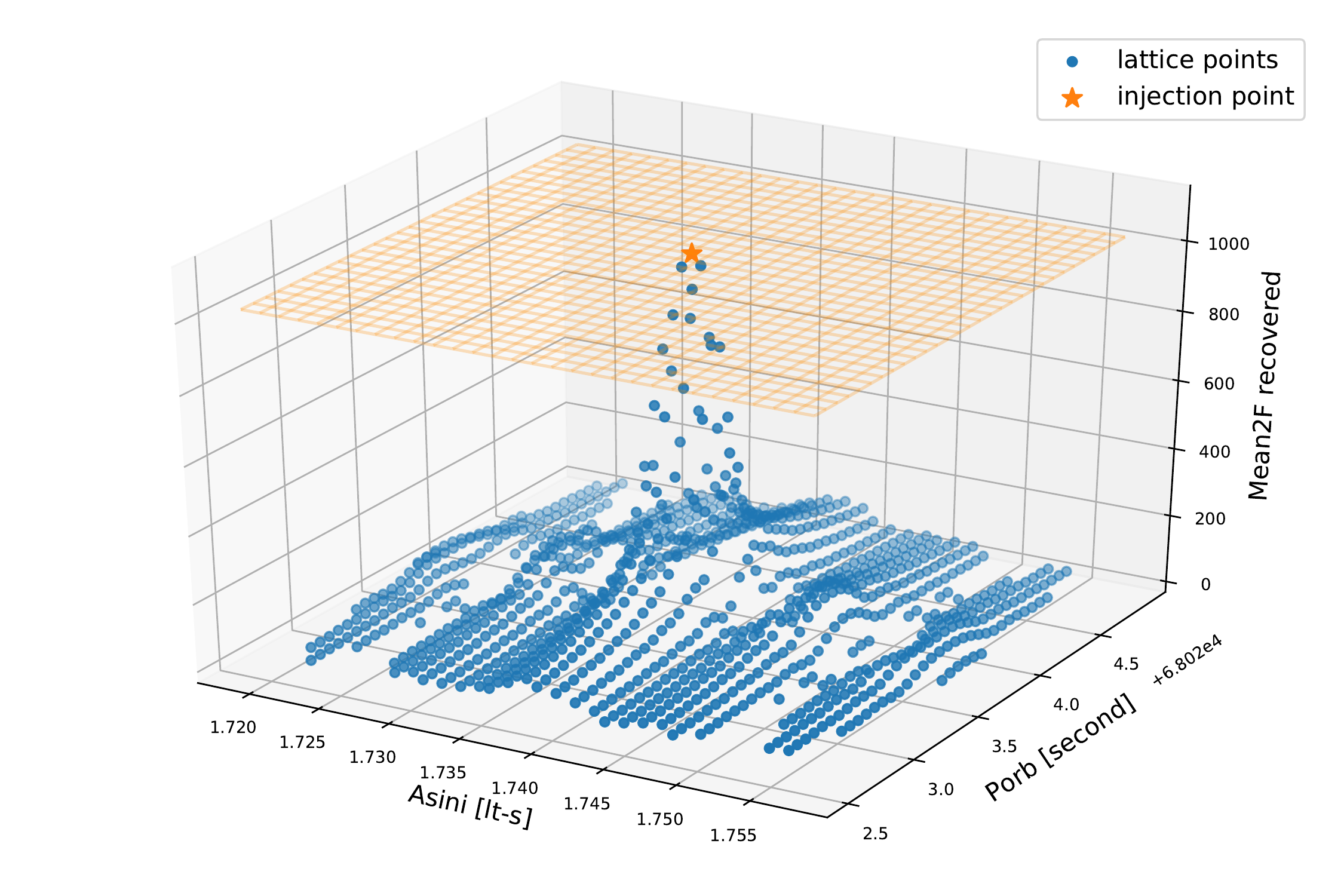}\\
  \includegraphics[clip,width=\columnwidth]{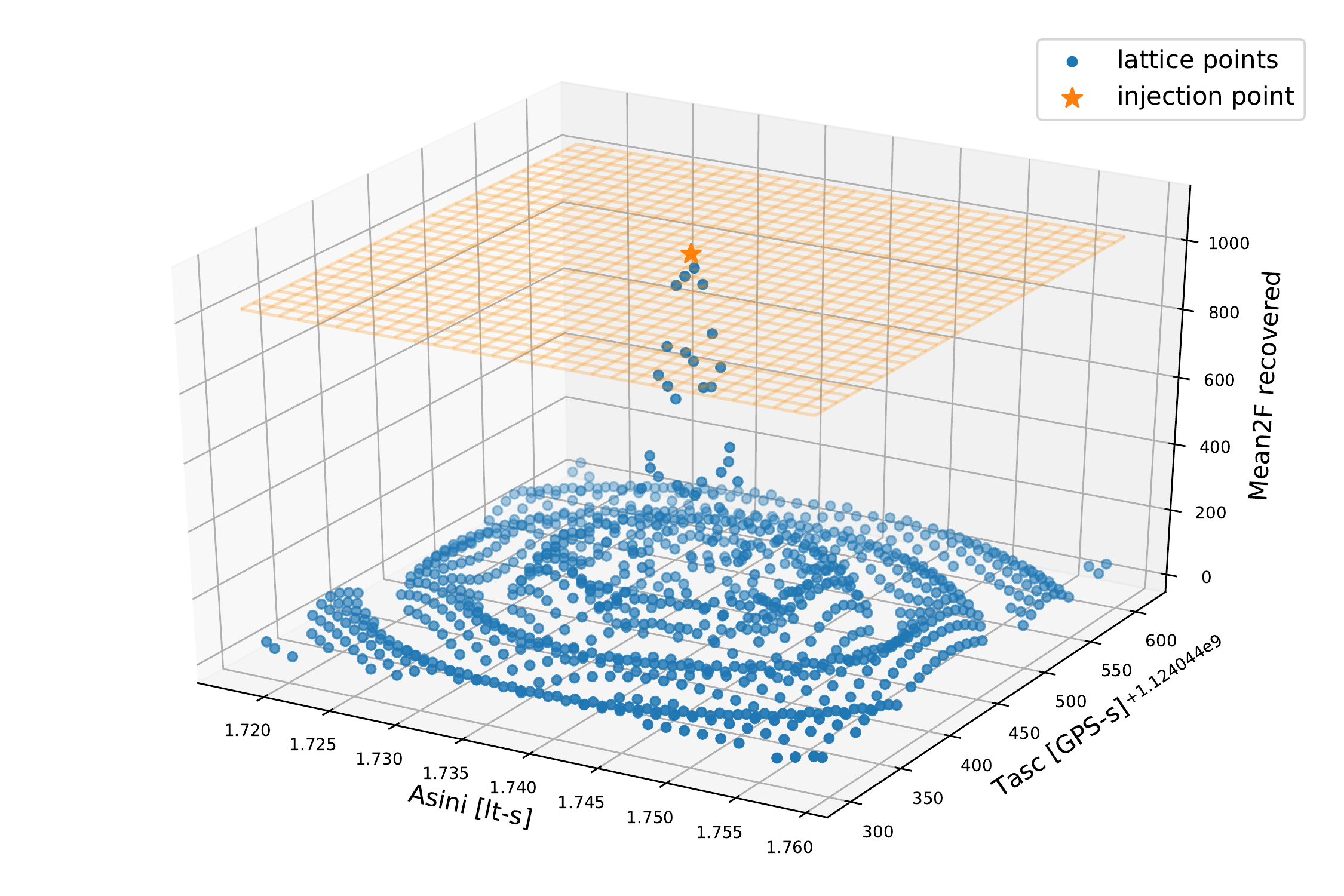}
  \includegraphics[clip,width=\columnwidth]{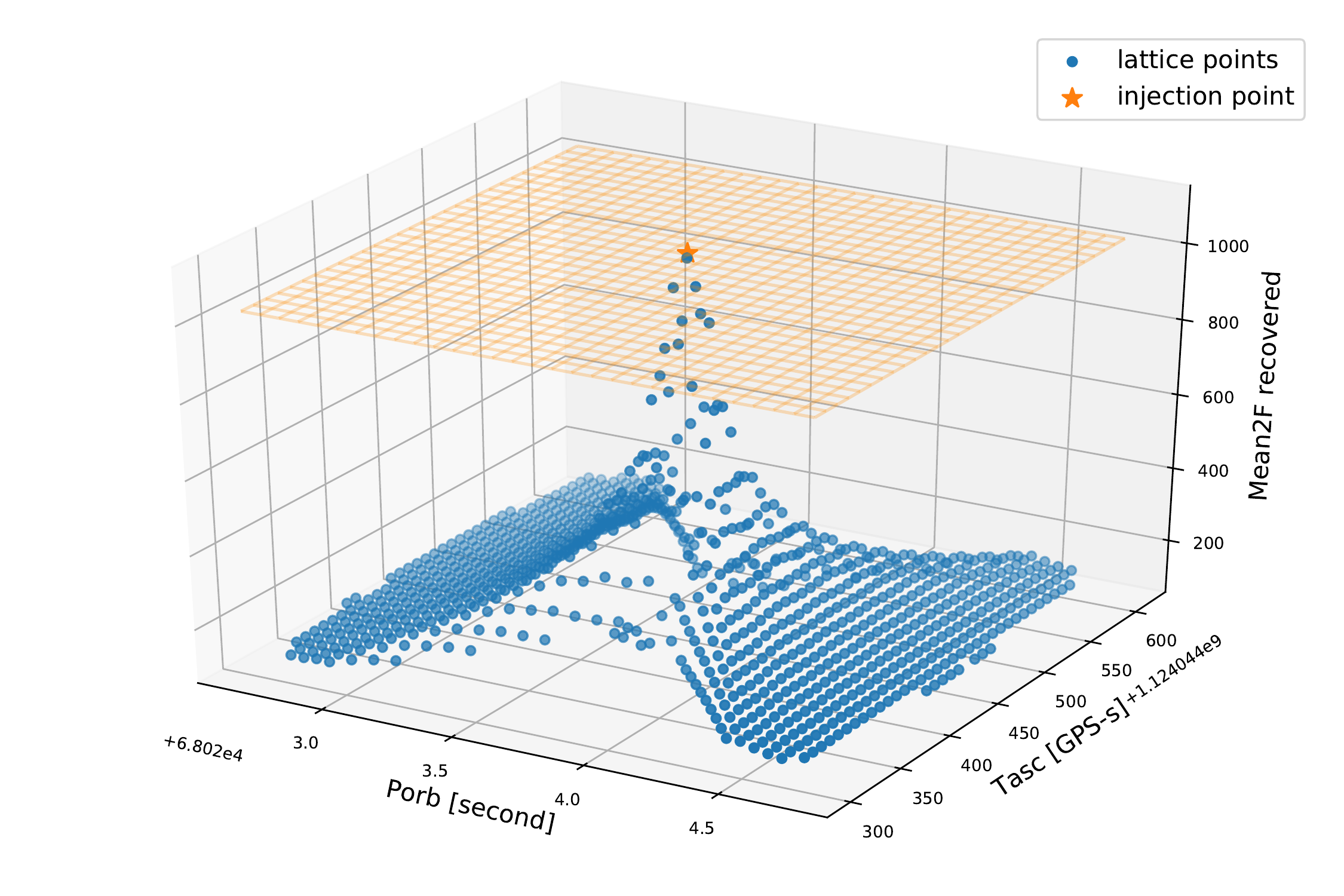}\\
  \caption{Illustration of 2D template-bank searches around a noiseless signal injection, with the respective
    two remaining search parameters fixed to the injected signal.
    The filled circles mark the placement of templates and their corresponding measured $\Fsco$-statistic
    values, while the star marks the signal injection point with its corresponding perfect-match
    $\Fsco$-statistic.
    The template bank was constructed for a maximum mismatch of $\mmax = 0.05$, with $\Nseg=120$ segments of
    $\Tseg=\SI{3}{\day}$.
    The horizontal mesh grid denotes the $\Fsco$-value corresponding to the maximum-mismatch criterion
    Eq.~\eqref{eqn:mismatch_criteria} relative to the injected signal power.
  }
  \label{fig:templates_2param}
\end{figure*}
The 2D searches are performed along all six two-parameter combinations out of the four, with the remaining two
parameters fixed to the injected signal location, with one example shown in Fig.~\ref{fig:templates_2param}.

These results illustrate the maximum-mismatch criterion of Eq.~\ref{eqn:mismatch_criteria} being satisfied, as
well as placing only one template in the ``vicinity'' $<\mmax$ of the signal as desired for an
efficient template bank.

\subsubsection{Testing 3D and 4D lattice tilings}
\label{sec:testing-3d-4d}

\begin{figure*}[htbp]
  \raggedright \hspace*{\columnwidth}\\
  \includegraphics[clip,width=1.1\columnwidth]{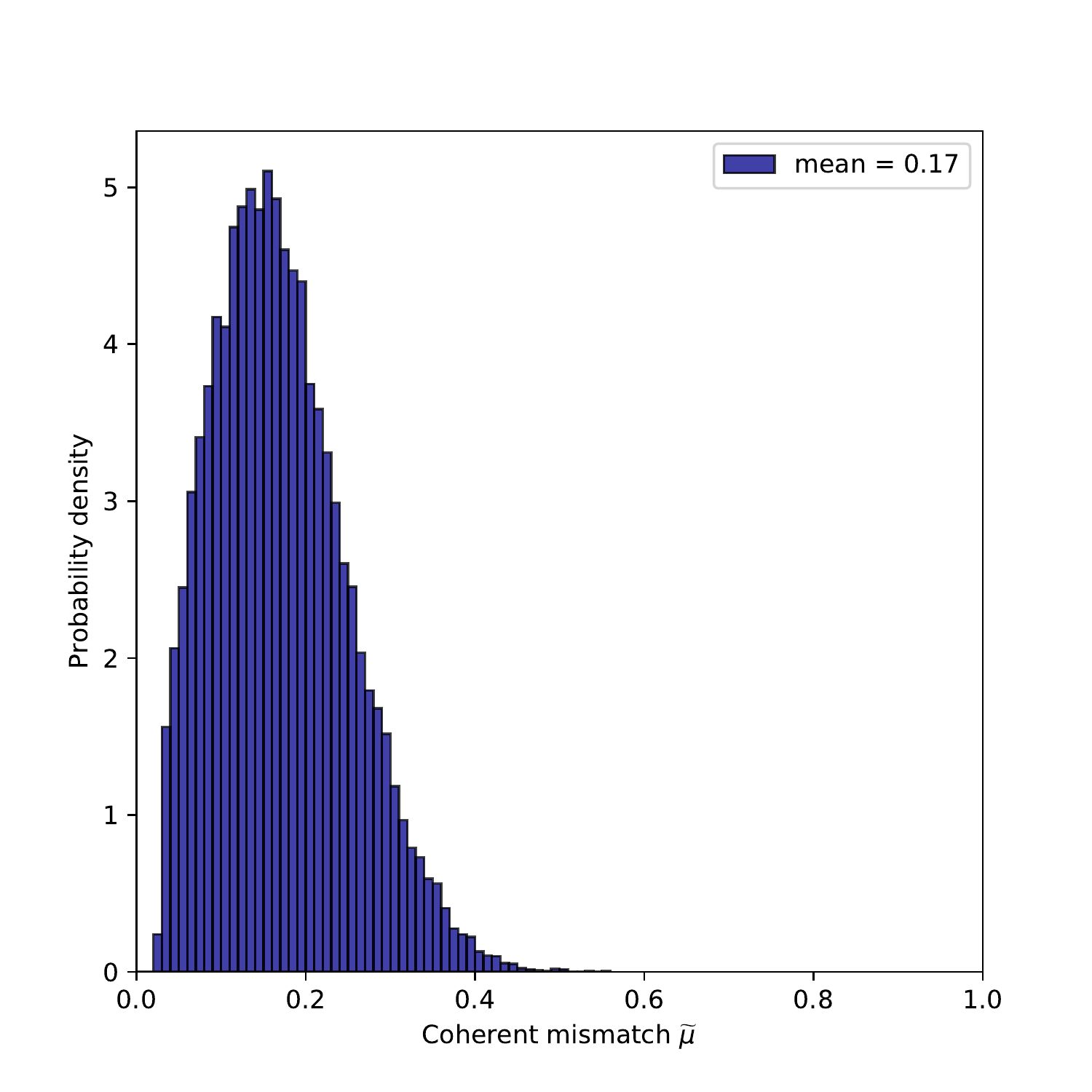}\hspace*{-0.5cm}
  \includegraphics[clip,width=1.1\columnwidth]{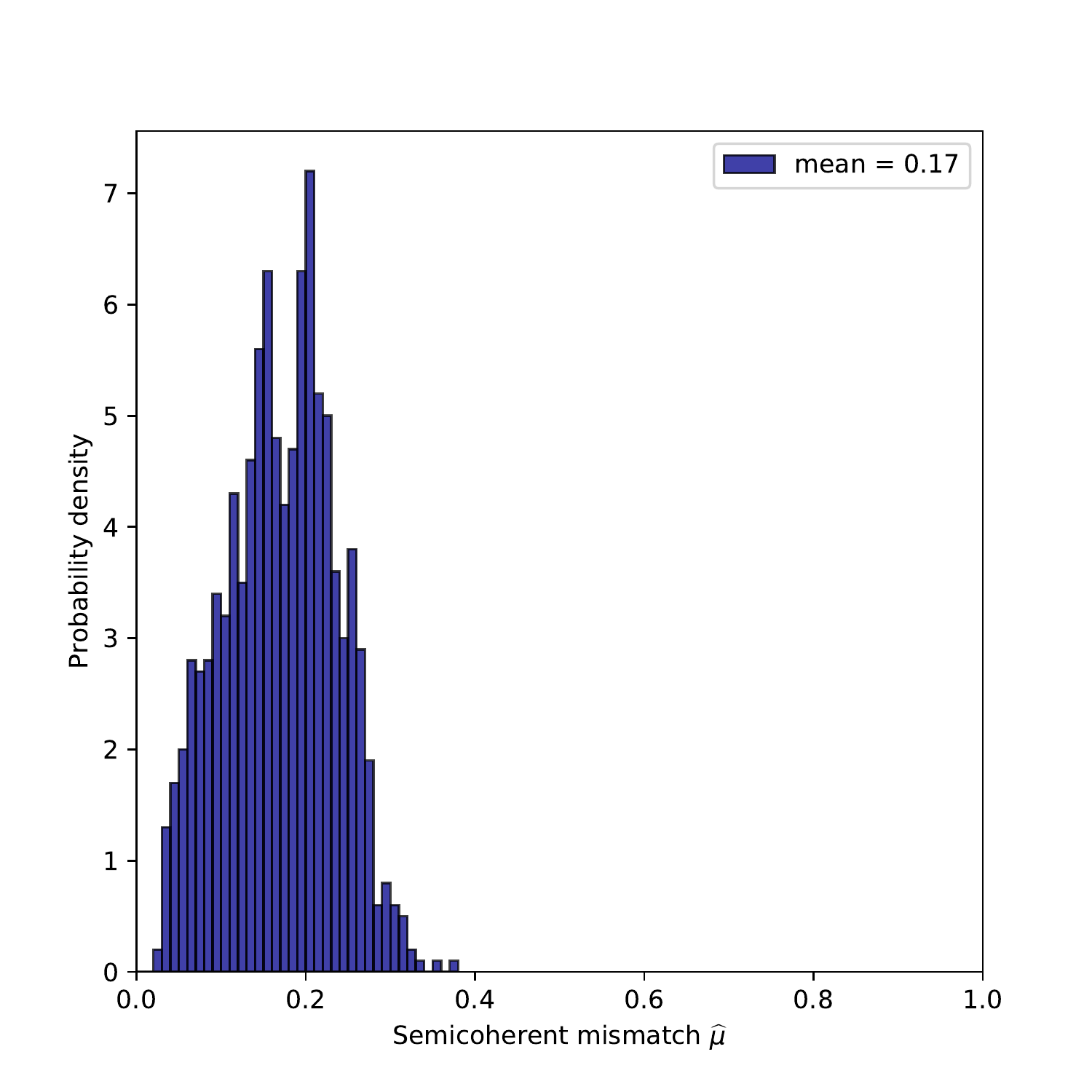}\hspace*{-0.5cm}

  \caption{Distribution of coherent per-segment mismatches $\coh{\mis}_0$ (left plot) and semi-coherent
    mismatches $\sco{\mis}_0$ (right plot), obtained from \num{1000} simulated 3D searches over a small box
    in $f$, $\asini$ and $t\asc$ around the injected signals (with $\Porb$ fixed at its injection value),
    with parameters drawn randomly from the test range $\Dop_0$ defined in
    Table.~\ref{tab:search_param_spaces}.
    The template bank was constructed for a maximum mismatch of $\mmax = 0.5$, with $\Nseg=30$ segments of
    $\Tseg=\SI{1}{\day}$.
  }
  \label{fig:mismatchHist_3param}
\end{figure*}
\begin{figure*}[htbp]
  \raggedright \hspace*{\columnwidth}\\
  \includegraphics[clip,width=1.1\columnwidth]{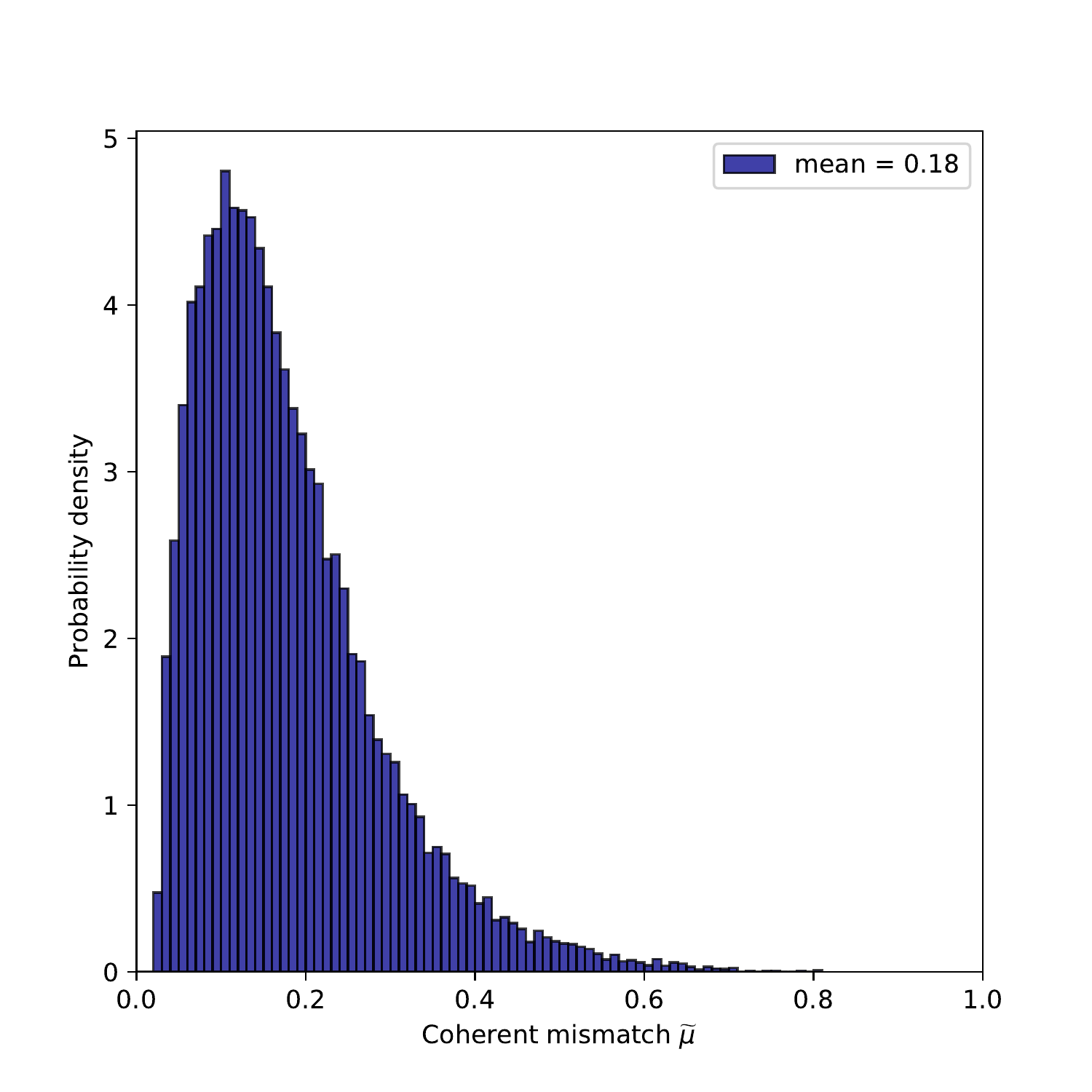}\hspace*{-0.5cm}
  \includegraphics[clip,width=1.1\columnwidth]{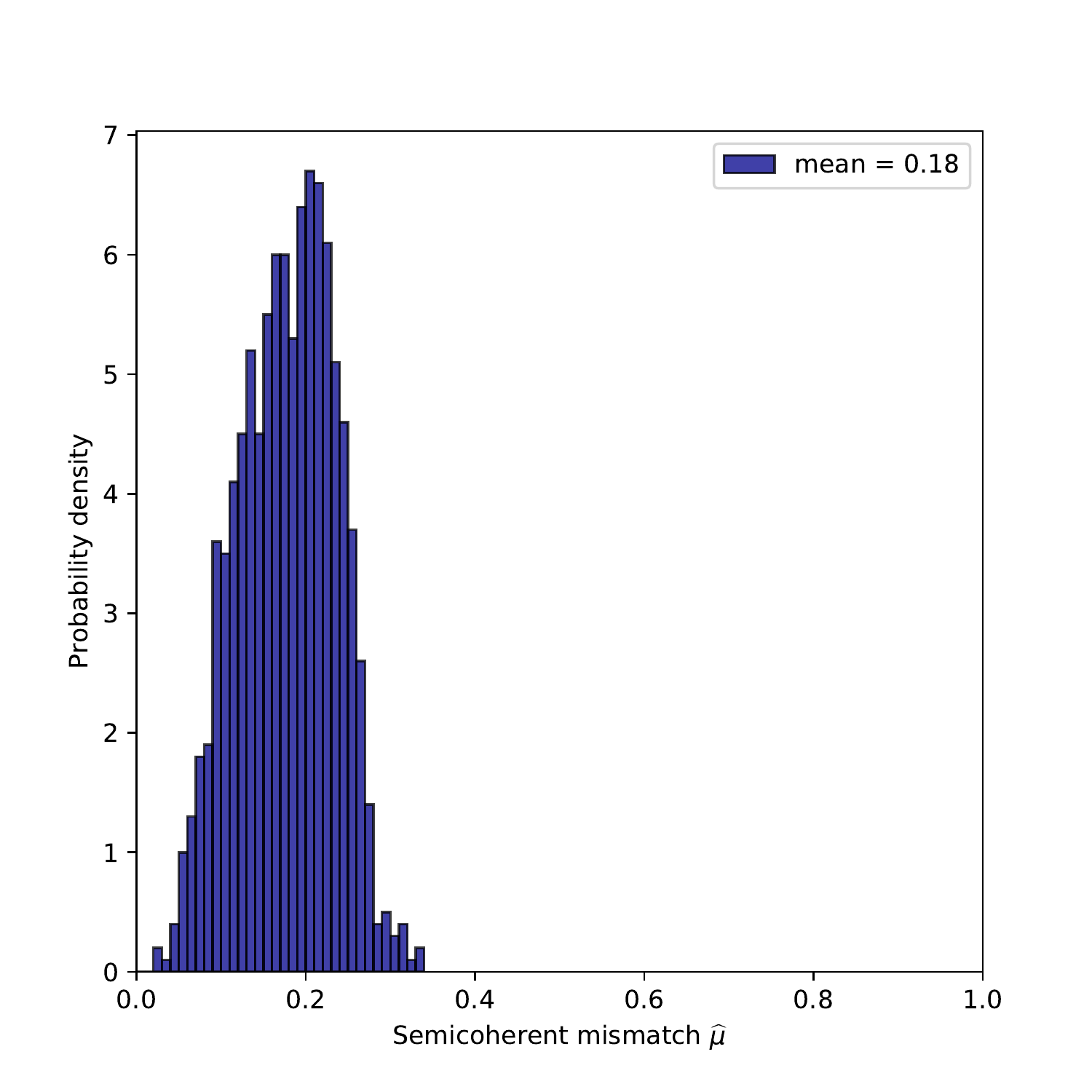}\hspace*{-0.5cm}

  \caption{Distribution of coherent per-segment mismatches $\coh{\mis}_0$ (left plot) and semi-coherent
    mismatches $\sco{\mis}_0$ (right plot), obtained from \num{1000} simulated 4D searches
    over a small box in $f, \asini, t\asc$ and $\Porb$ around the injected signals, with parameters drawn
    randomly from the test range $\Dop_0$ defined in Table.~\ref{tab:search_param_spaces}.
    The template bank was constructed for a maximum mismatch of $\mmax = 0.5$, with $\Nseg=30$ segments of
    $\Tseg=\SI{1}{\day}$.
  }
  \label{fig:mismatchHist_4param}
\end{figure*}
\begin{figure*}[htbp]
  \raggedright \hspace*{\columnwidth}\\
  \includegraphics[clip,width=1.1\columnwidth]{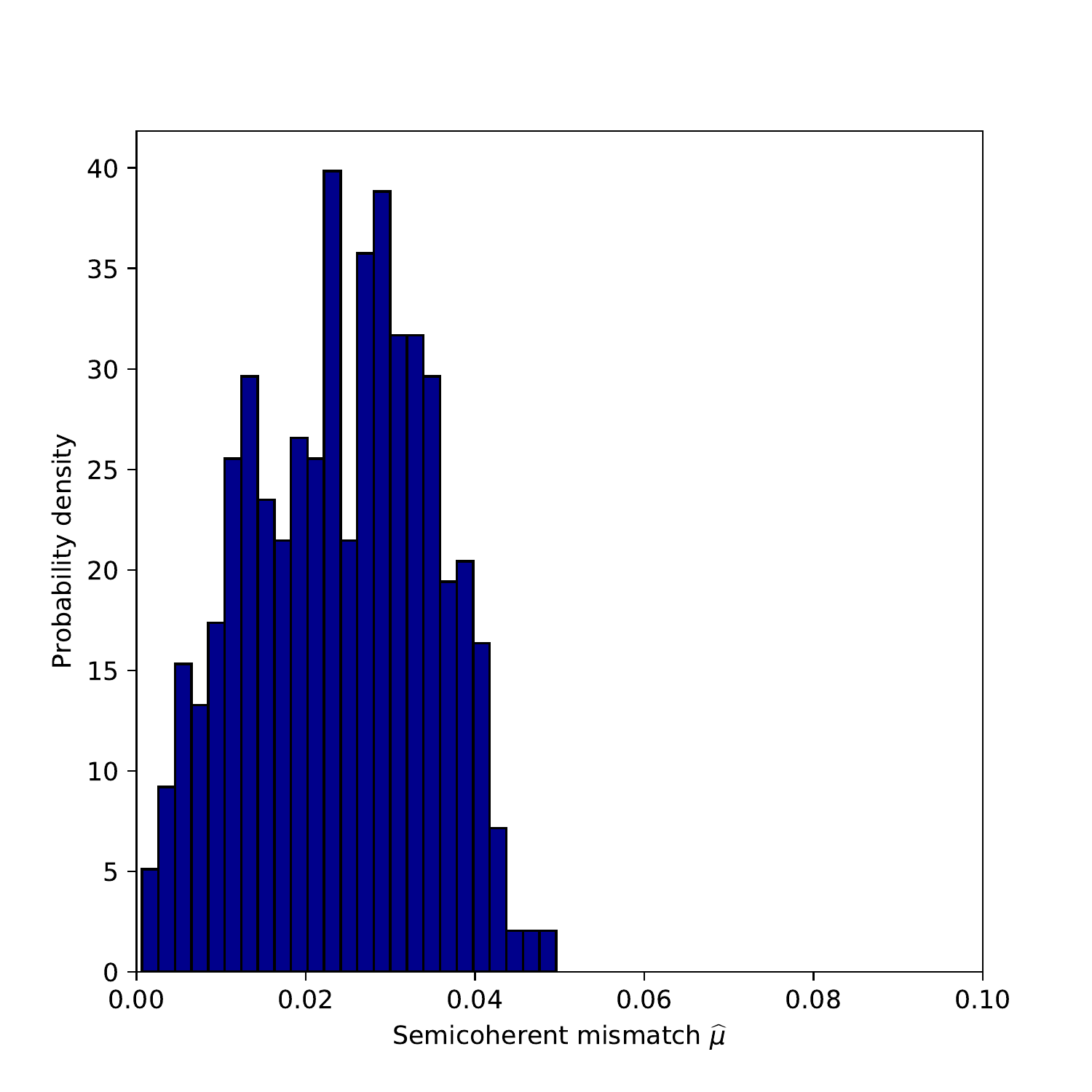}\hspace*{-0.5cm}
  \includegraphics[clip,width=1.1\columnwidth]{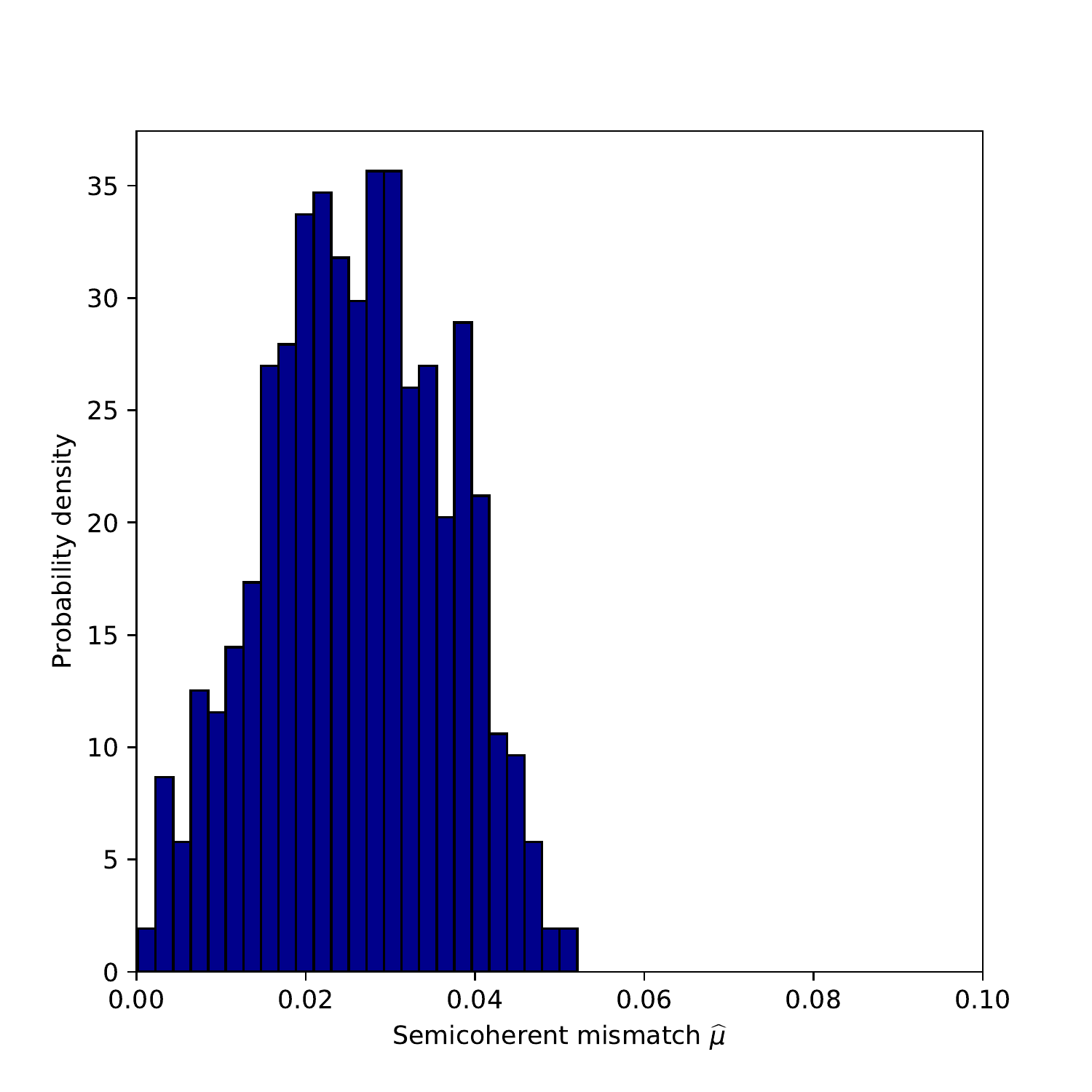}\hspace*{-0.5cm}

  \caption{Distribution of semi-coherent mismatches $\sco{\mis}_0$ obtained from \num{500} simulated 4D 
    searches (left plot) over a small box in $f$, $\asini$ and $t\asc$ around the injected signals (with 
    $\Porb$ fixed at its injection value); and distribution of \num{500} simulated 4D searches (right plot)
    over a small box in $f, \asini, t\asc$ and $\Porb$ around the injected signals, with parameters drawn 
    randomly from the test range $\Dop_0$ defined in Table.~\ref{tab:search_param_spaces}. The template 
    bank was constructed for a maximum mismatch of $\mmax = 0.05$, with $\Nseg=120$ segments of 
    $\Tseg=\SI{3}{\day}$.
  }
  \label{fig:mismatchHist_smallmumax}
\end{figure*}

\begin{table*}[htbp]
\begin{center}
\begin{tabular}{ |l|c|c|c|c|c| }
 \hline
Search space $\Dop$ & $f\,[\si{Hz}]$ & $\asini\,[\si{ls}]$ & $\Porb\,[\si{s}]$ & $t\asc\,[\si{GPS\,s}]$ & Reference(s)/comment(s) \\
  \hline
  $\Dop_0$     & 10--700   &  0.3--3.5  & 68023.7 $\pm$ 0.2      & 1124044455.0 $\pm$ 1000 &  \BinaryWeave{} test range \\
  \hline
  {$\Dop_1$}   &   20--500  & 1.26--1.62 & 68023.70496 $\pm$ 0.0432 &  897753994 $\pm$ 100 & \citet{BinaryWeave:method} \\
  {$\Dop_2$}   &   60--650  & 1.45--3.25 & 68023.86048 $\pm$ 0.0432 &  974416624 $\pm$  50 & \citet{ScoX1O2LVC:prd2019} \\
  {$\Dop_3$}   &   40--180  & 1.45--3.25 & 68023.86    $\pm$ 0.12   & 1178556229 $\pm$ 417 & \citet{ScoX1O2_AEI:apjl2021} \\
  \hline
  {$\Dop_4$}   &  600--700  &            &                          &                      &                               \\
  {$\Dop_5$}   & 1000--1100 &            &                          &                      &                               \\
  {$\Dop_6$}   & 1400--1500 & 1.45--3.25 & 68023.70496 $\pm$ 0.0432 &  974416624 $\pm$ 100 & different ranges in frequency \\
  {$\Dop_7$}   &   20--250  &            &                          &                      & with broad range in $\asini$  \\
  {$\Dop_8$}   &   20--1000 &            &                          &                      &                               \\
  {$\Dop_9$}   &   20--1500 &            &                          &                      &                               \\
  \hline
  {$\Dop_{10}$}  &  600--700  &            &                          &                      &                               \\
  {$\Dop_{11}$}  & 1000--1100 &            &                          &                      &                               \\
  {$\Dop_{12}$}  & 1400--1500 & 1.40--1.50 & 68023.70496 $\pm$ 0.0432 &  974416624 $\pm$ 100 & different ranges in frequency \\
  {$\Dop_{13}$}  &   20--500  &            &                          &                      & with narrow range in $\asini$ \\
  {$\Dop_{14}$}  &   20--1000 &            &                          &                      &                               \\
  {$\Dop_{15}$}  &   20--1500 &            &                          &                      &                               \\
  \hline
  {$\Dop_{16}$}  &  600--700  &            &                          &                      &                               \\
  {$\Dop_{17}$}  & 1000--1100 &            &                          &                      &                               \\
  {$\Dop_{18}$}  & 1400--1500 & 1.44--1.45 & 68023.70496 $\pm$ 0.0432 &  974416624 $\pm$ 100 & different ranges in frequency \\
  {$\Dop_{19}$}  &   20--500  &            &                          &                      & with well-constrained $\asini$\\
  {$\Dop_{20}$}  &   20--1000 &            &                          &                      &                               \\
  {$\Dop_{21}$}  &   20--1500 &            &                          &                      &                               \\
 \hline
\end{tabular}
\end{center}
\caption{Different parameter space search regions considered for \Sco{}.
  $\Dop_0$ has been used in this study as a test range for various Monte-Carlo tests of \BinaryWeave{}.
  $\Dop_{1-3}$ represent observational constraints considered in recent CW searches and studies.
  In addition, various combinations of parameter-ranges are considered, $\Dop_{4-21}$, in order to
  explore the impact of improved observation constraints and reduced search ranges.
  }
\label{tab:search_param_spaces}
\end{table*}

Next we test the template-bank performance for the four possible combinations of \emph{three} search
parameters (3D searches) with the fourth one fixed to the signal injection parameter, as well as 4D searches
over all four parameters $\{f,\asini,\Porb,t\asc\}$.
We perform several sets of simulations, using $\sim \Ord{100 - 1000}$ injections each, using varying search
setups and maximum mismatch values $\mmax$, in order to obtain the resulting mismatch distribution of the
template bank.

The injected signal parameters are randomly drawn from the \emph{test range} $\Dop_0$
(cf.~\ref{tab:search_param_spaces}), namely $f\in[10, 700]\,\si{Hz}$ and binary parameter ranges wider than
the \Sco{} constraints, namely $\asini\in[0.3-3.5]\,\si{ls}$,
$\Porb = \SI{68023.7 \pm 0.2}{\second}$ and $t\asc = \SI{1124044455.0 \pm 1000}{GPS\,\second}$.

Figure~\ref{fig:mismatchHist_3param} shows an example for the mismatch distributions of coherent
and semi-coherent mismatches obtained for a set of \num{1000} injections and subsequent 3D searches in a small
box around the injection in $f$, $\asini$ and $t\asc$, with $\Porb$ fixed to the injected value.
Figure~\ref{fig:mismatchHist_4param} presents a corresponding example for the mismatch distributions obtained
from \num{1000} 4D box searches around the injected signals.

We see that the means of the coherent and semicoherent mismatch distributions are
$\avg{\coh{\mis}} \approx \avg{\sco{\mis}} \approx 0.17-0.18$, and the highest observed semicoherent mismatch in the 3D
case is $\max{\sco{\mis}_0} \approx 0.4$, while in the 4D case it is
$\max{\sco{\mis}_0} \approx 0.35$.
This is smaller than the imposed maximum mismatch of $\mmax = 0.5$, which is a common feature of the
quadratic approximation Eq.~\eqref{eqn:metric_mismatch} underlying the metric, namely the measured
mismatch values $\mis_0$ tend to increasingly fall behind the predicted metric mismatch values with increasing
mismatch \citep[e.g., see ][]{Reinhard_MultiDetFstat,WettePrix_2013PRD,allen_spherical_2019}. Thus,
in addition, we also test the metric mismatch implementations for small mismatch value $\mmax = 0.05$
which is compareable to the realistic search setups relevant for Sco X-1 (discussed in details in
Section~\ref{sec:sensitivity_estimate}). We see a good agreement for both 3D and 4D template banks with
such small $\mmax$ values as shown in Figure~\ref{fig:mismatchHist_smallmumax}.

\subsection{Required computing resources}
\label{banksize_computing_resources}

\subsubsection{Number of templates}
\label{num_templates}

As discussed in Sec.~\ref{subsec:template_mismatch}, the \emph{bulk} template count for a parameter space
$\Dop$ (not counting any extra templates required for boundary padding of $\partial\Dop$) is given by
Eq.~\ref{eqn:number_templates}.

Using the metric expressions in Eq.~\eqref{eqn:semicoh_phase_metric}, this can be
evaluated explicitly \cite{BinaryWeave:method} and the bulk template count for 3D searches over
$\{f,\asini,t\asc\}$ is found as
\begin{equation}
  \label{eqn:number_templates_3D}
  \begin{split}
    \sco{\N}_{3\dims} = \frac{\theta_{3}}{\mmax^{3/2}} \frac{\pi^3 \Tseg}{\sqrt{27}}\,\Omega\,
    (f^3_{\max}-f^3_{\min})\,(\asini^{2}{}_{,\max} - \asini^2{}_{,\min}) \\
    \times (t\asc{}_{,\max} - t\asc{}_{,\min}),
  \end{split}
\end{equation}
while for a 4D template bank over $\{f, \asini, \Porb, t\asc\}$ one finds
\begin{equation}
\label{eqn:number_templates_4D}
    \begin{split}
      \sco{\N}_{4\dims} = \frac{\theta_{4}}{\mmax^{2}} \frac{\pi^4 \gamma \Tseg^2}{36\sqrt{2}}
	(f_{\max}^{4} - f_{\min}^{4}) (\asini{}_{,\max}^{3} - \asini{}_{,\min}^{3}) \\
	\times (\Omega_{\max}^{2} - \Omega_{\min}^{2}) (t\asc{}_{,\max} - t\asc{}_{,\min}),
    \end{split}
\end{equation}
where the coordinate ranges are $\lambda^i\in[\lambda^i_{\min},\lambda^i_{\max}]$, and $\gamma$ is the
semi-coherent \emph{refinement factor} associated with the $\Porb$ (i.e., $\Omega$), given by
\begin{equation}
\label{eqn:refinement_factor}
	\gamma = \sqrt{1 + 12\frac{(\av{\Delta}\ma^2 - \av{\Delta\ma^2})}{\Tseg^2}}.
\end{equation}
The refinement factor evaluates to $\gamma=\Nseg$ in the case of segments without gaps.
We can use these theoretical expressions to test against the actual number of templates generated by the
\BinaryWeave{} code, which includes boundary padding not accounted for in the above theoretical expressions.

\begin{figure*}[htbp]
  \raggedright \hspace*{\columnwidth}\\
  \includegraphics[clip,width=\columnwidth]{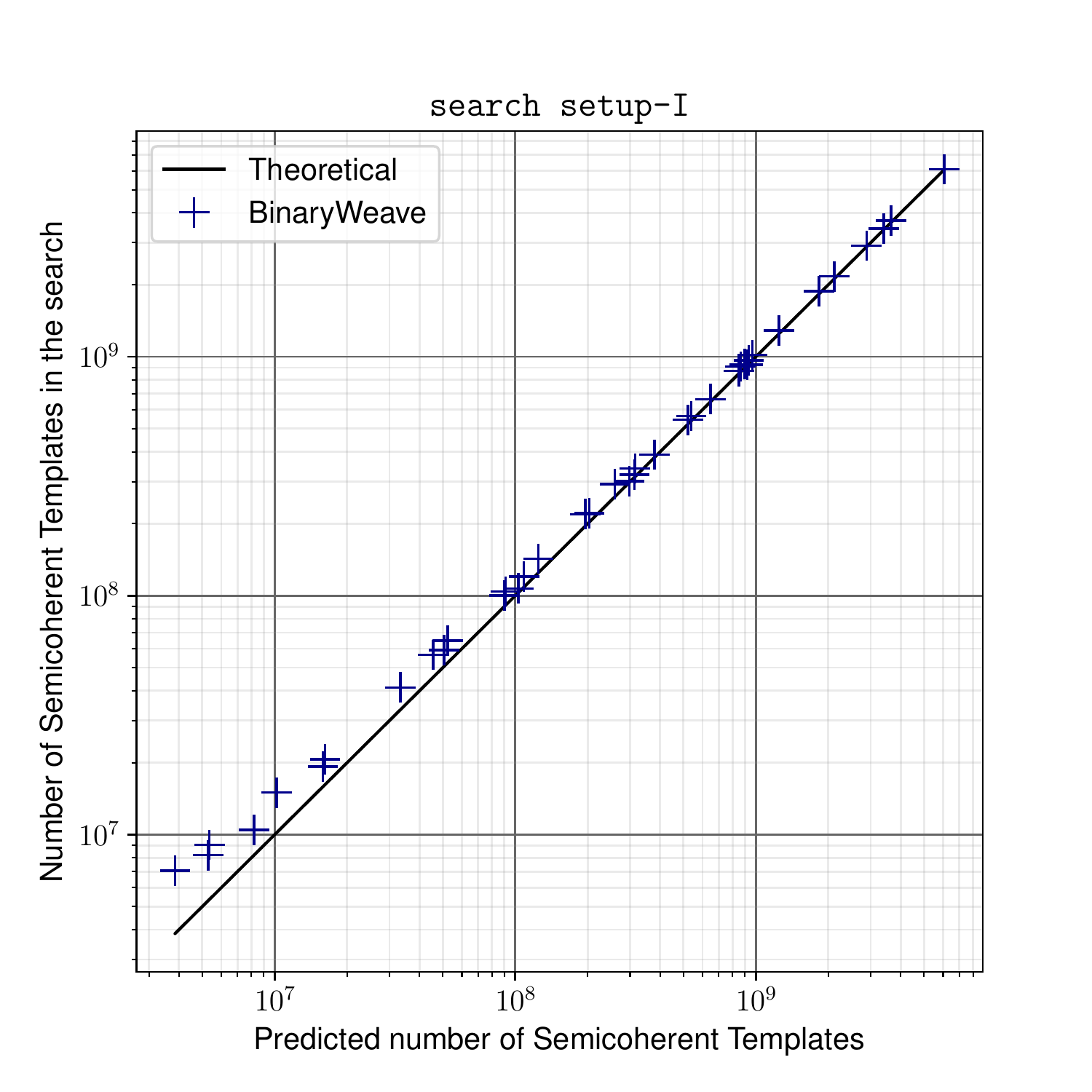}
  \includegraphics[clip,width=\columnwidth]{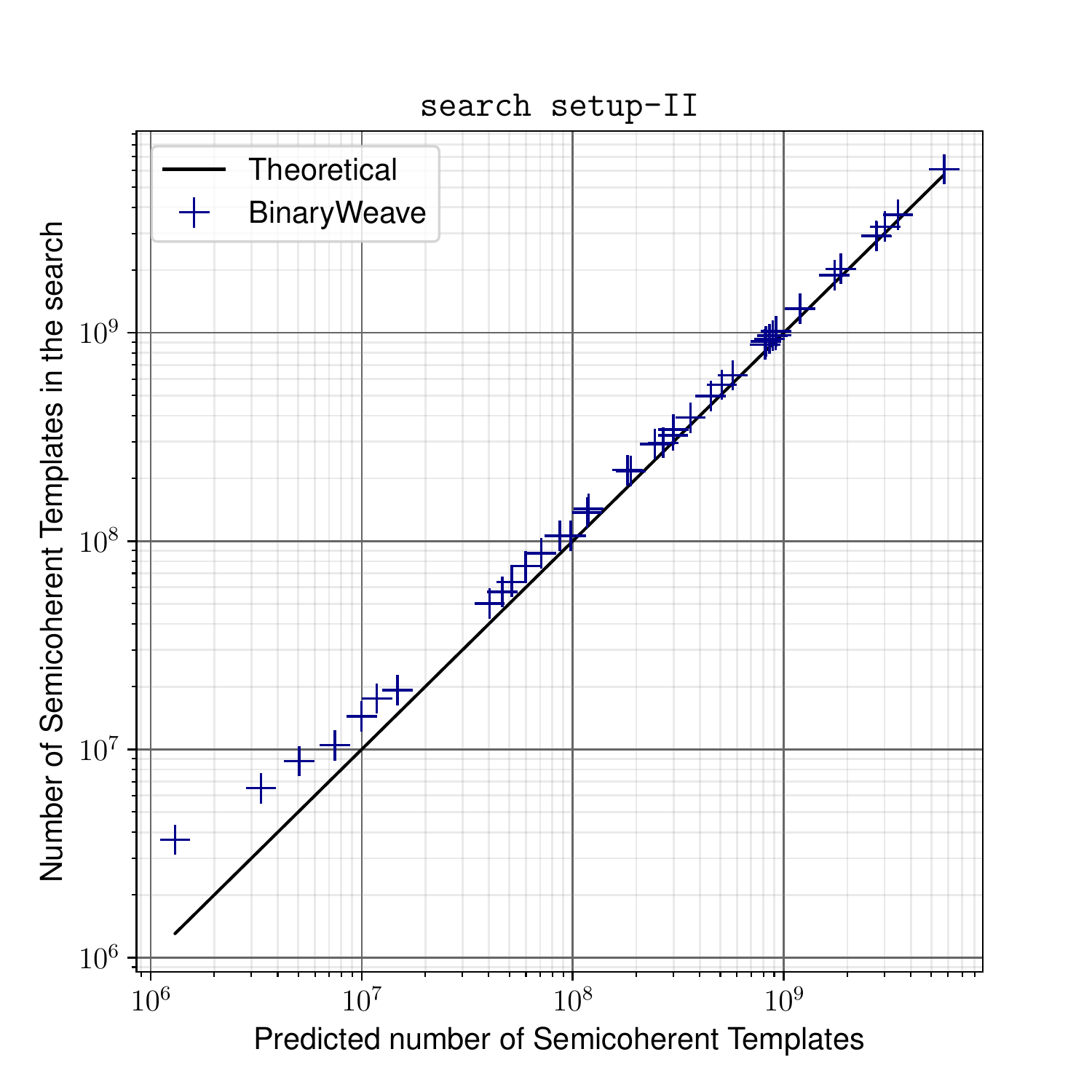}
  \caption{Number of semicoherent templates $\sco{\N}$ constructed by \BinaryWeave{} versus with the theoretical
    \emph{bulk} predictions of Eq.~\eqref{eqn:number_templates_4D}.
    Each point `$+$' corresponds to a simulated 4D-box search around a randomly chosen parameter-space
    location in $\{f,\asini\}\in\Dop_0$ (cf.\ Table~\ref{tab:search_param_spaces} using either \setup{I}
    (left plot) or \setup{II} (right plot) defined in Table~\ref{tab:search_setup}).
  }
  \label{fig:compare_num_tmplt}
\end{figure*}

In the following we consider two example search setups (cf.\ Table~\ref{tab:search_setup}), namely \setup{I}
with $\Nseg=180$ segments of duration $\Tseg=\SI{1}{\day}$ and a maximum mismatch of $\mmax=0.031$, and
\setup{II} with $\Nseg=120$ segments of $\Tseg=\SI{3}{\day}$ and maximum mismatch $\mmax=0.056$.

We generate a \BinaryWeave{} template bank for a small \emph{box} around a randomly-chosen point in
$f$ and $\asini$, drawn from the test range $\Dop_0$ of Table.~\ref{tab:search_param_spaces}.
The box consist of $\Ord{\num{e5}}$ frequency bins and a metric bounding-box extent $D\lambda^i$
(cf.~Eq.~\eqref{eq:5}) along each binary-orbital parameter dimension.
This is repeated 40 times, in order to obtain a representative sampling over a wide range of search
parameters, and the resulting \BinaryWeave{} template counts are compared to the theoretical
predictions of \eqref{eqn:number_templates_4D}, shown in Fig.~\ref{fig:compare_num_tmplt}.

We see that there is generally good agreement in the template counts, with the real template counts exceeding
the theoretical \emph{bulk} predictions by factors up to $2-3$ at low template counts, with increasingly
good agreement at higher template counts. The template counts exceeds only at the lowest frequency regime
($\leq 50$ Hz) by a factor of $\sim 2-3$, whereas agrees within $10 \%$ at intermediate frequency ($\sim 
200$ Hz) and $< 5 \%$ at higher frequency ($\sim 500$ Hz). 
This effect is expected from the extra padding required to fully cover the parameter-space boundaries
$\partial\Dop$, which decreases in relative importance for increasing total template counts (i.e., boundary
effects are less important for template spacings that are small compared to the parameter-space extents).

\subsubsection{Computing cost and memory usage}
\label{cpu_runtime}

A detailed computing-cost (and memory) model exists for the semi-coherent \Weave{} implementation
\cite{IsolatedWeave} as well as for the underlying coherent $\F$-statistic implementation
\cite{prix:Fstat_timing}.
There are two different $\F$-statistic algorithms available, the \emph{resampling FFT} algorithm
(originally described in \cite{Fstat:JKS}), and the so-called \emph{demodulation} algorithm
introduced in \cite{williams_efficient_2000,prix_f-statistic_2010}. Because the resampling $\F$-statistic
is substantially faster (i.e., $\Ord{100-1000}$) for large numbers of frequency bins (i.e., $\Ord{10^5}$)
and SFTs, which is the relevant regime for the wide parameter-space search considered here, we will
exclusively consider this algorithm for the following discussion of the \Sco{} computing cost
\footnote{A GPU port of the resampling $\F$-statistic~\cite{DunnEtAl2022:GPrUImpFsCnGrvWS}, which
yields speedup factors of $\Ord{10-100}$, was developed after this study had been performed. A practical
application of the GPU resampling $\F$-statistic with \Weave{} can be found in \cite{Wette_etal_2021PRD}.}.

We performed the \BinaryWeave{} tests and simulations on the LIGO Data Analysis System (LDAS) computing
cluster at the LIGO Hanford Observatory, containing a combination of \texttt{2.4GHz Xeon E5-2630v3},
\texttt{2.2GHz Xeon E5-2650v4}, \texttt{3.5GHz Xeon E3-1240v5} and \texttt{3.0GHz Xeon Gold 6136} CPUs.
We find the resulting semi-coherent timing coefficients measured on this hardware are essentially the same as
given in Table.~III of \cite{IsolatedWeave}, while the effective (resampling-FFT) $\F$-statistic time per
template and detector is observed to fall in the range $\tau_\F\eff \approx (3.8-4.3)\times\SI{e-7}{\second}$,
consistent with the numbers obtained in \cite{IsolatedWeave}.

We measure the CPU run-time per template $\Cost\tmpl$ and the maximum memory usage of \BinaryWeave{} for the
80 box searches (two sets of 40 box searches each for setup-I and setup-II) described in the previous section
(see Fig.~\ref{fig:compare_num_tmplt}).
The maximum memory usage over all search boxes is found as $\sim\SI{2.2}{GB}$, well below all-sky \Weave{}
numbers observed in \cite{2019arXiv190108998W}, due to the fact that \Sco{} has little refinement and we can
use a non-interpolating search setup, substantially alleviating memory requirements.

\begin{figure*}[htbp]
  \includegraphics[width=\columnwidth]{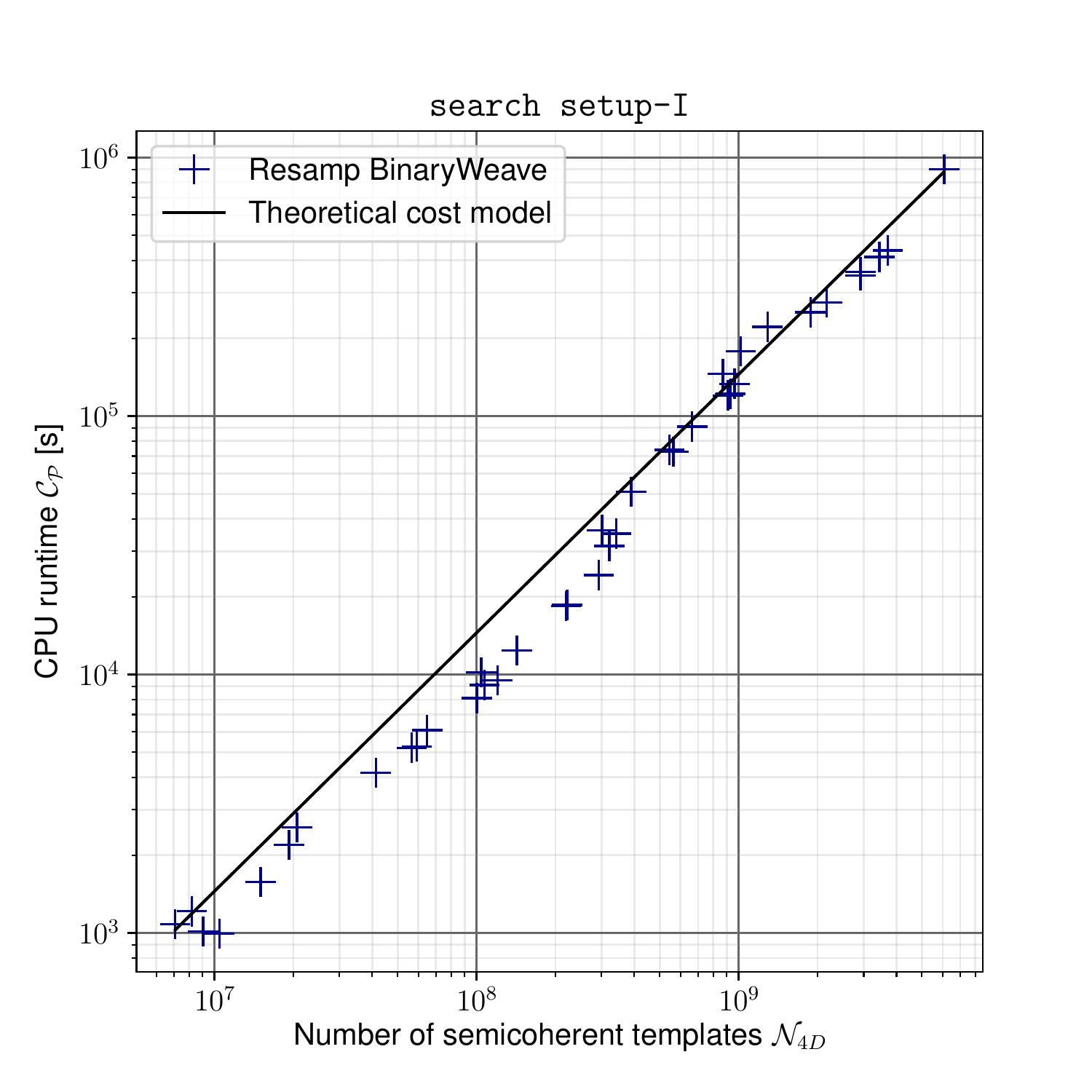}
  \includegraphics[width=\columnwidth]{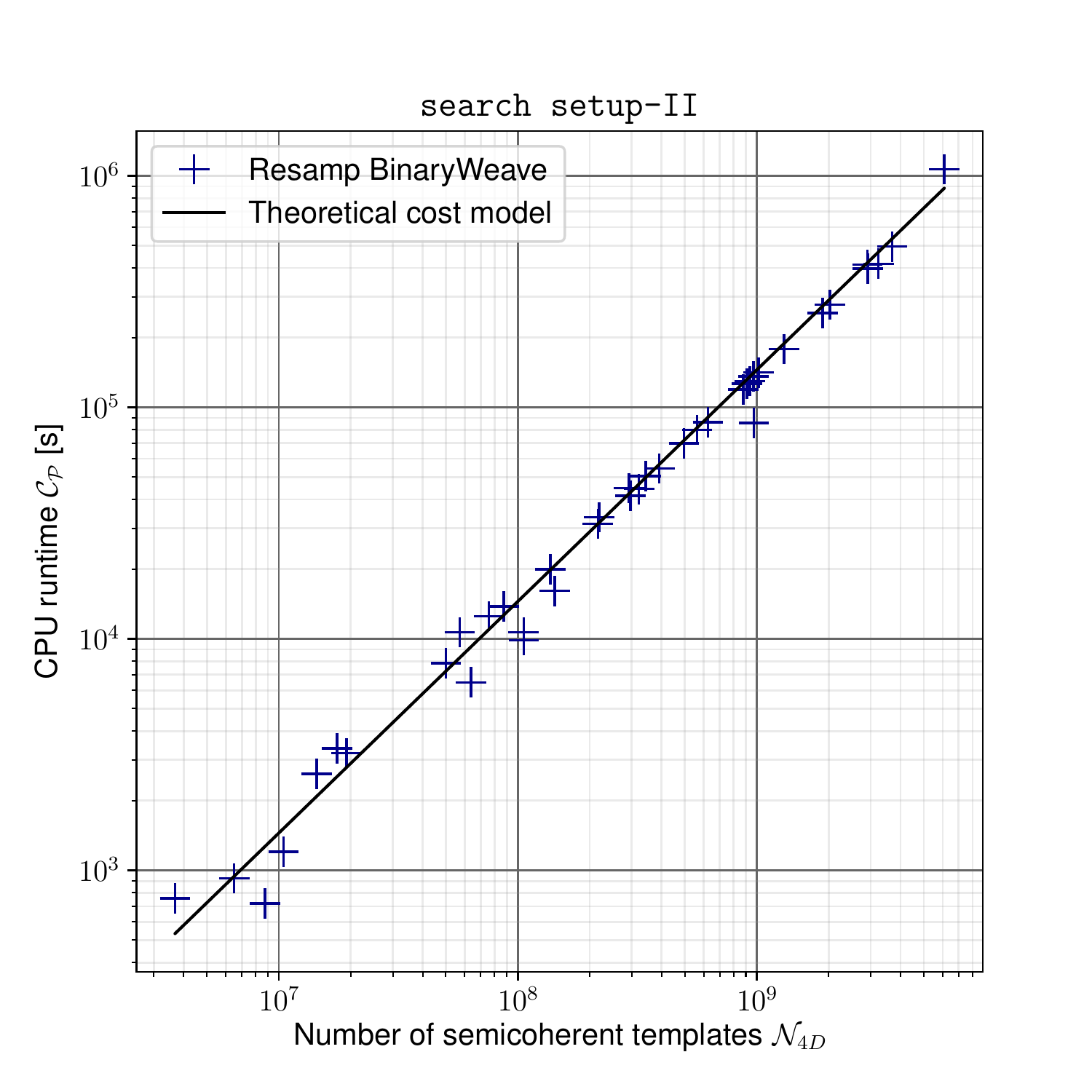}
  \caption{CPU run-time $\Cost_{\Dop}$ per search box as a function of the number of (semi-coherent)
    templates $\sco{\N}_{4\dims}$ for that box, for \setup{I} (left plot) and \setup{II} (right plot), defined
    in Table.~\ref{tab:search_setup}.
    The points '$+$' mark the measured \BinaryWeave{} run times, while the solid line indicates the
    effective cost model prediction of Eq.~\eqref{eq:8}, using an effective cost per template of
    $\Cost\tmpl=\SI{0.145}{ms}$.
  }
  \label{fig:cpu_runtime}
\end{figure*}

The runtime per template $\Cost\tmpl$ is found to be relatively constant over the search parameter space and
for the two search setups considered, namely
$\Cost\tmpl(\setup{I}) \approx \SI{0.12 +- 0.03}{\milli\second}$ and
$\Cost\tmpl(\setup{II}) \approx \SI{0.14 +- 0.03}{\milli\second}$.
Here we only consider the non-interpolating StackSlide method, in which the coherent segments and the
semi-coherent $\Fsco$-statistic share the same template grid and number of templates $\N$, i.e., $\coh{\N}=\sco{\N}$.
This implies that both the coherent and semi-coherent contributions to the total computing cost are
proportional to $\N$.
Therefore we can use a simplified \emph{effective} model for the \emph{total} computing cost $\Cost_{\Dop}$
over a search space $\Dop$ in the form
\begin{equation}
  \label{eq:8}
  \Cost_{\Dop} = \N_{\Dop}\,\Cost\tmpl,
\end{equation}
where $\N_{\Dop}$ is the total number of templates covering the parameter space $\Dop$.
Given the above timing measurements for the two setups, in the following we assume a (slightly conservative)
effective CPU time per template of $\Cost\tmpl = \SI{0.145}{ms}$.
This simplified effective cost model is plotted against the measured \BinaryWeave{} run times in
Fig.~\ref{fig:cpu_runtime}.

\section{Characterizing potential \Sco{} searches}
\label{sec:sensitivity_estimate}

\subsection{Sensitivity for different search setups}
\label{sec:pred-sens-diff}

The sensitivity of a search is typically characterized by the weakest signal amplitude $h_{\pfa}^{\pdet}$
detectable at a false-alarm probability $\pfa$ with detection probability (or ``confidence level'') $\pdet$.
While this is astrophysically informative, for a given search \emph{method} it is often more instructive
\cite{dreissigacker_fast_2018} to use the \emph{sensitivity depth} $\depth_{\pfa}^{\pdet}$ instead, defined as
\begin{equation}
  \label{eqn:sens_depth}
  \depth_{\pfa}^{\pdet} \equiv \frac{\sqrt{\Sn}}{h_{\pfa}^{\pdet}},
\end{equation}
which characterizes the sensitivity of a method independently of the noise floor (i.e., power spectral
density) $\Sn$.

As discussed in \cite{wette_estimating_2012,dreissigacker_fast_2018}, the sensitivity of a semi-coherent
StackSlide $\Fsco$-statistic search can be estimated quite accurately (to better than $\sim\SI{10}{\percent}$)
given the total amount of data used, the number $\Nseg$ of semi-coherent segments and the mismatch
distribution of the template bank. This algorithm is implemented in the \textsc{OctApps}
\cite{wette_octapps_2018} function \texttt{SensitivityDepthStackSlide()}.

For each search setup listed in Table.~\ref{tab:search_setup} we obtain the mismatch distribution empirically
by injection-recovery Monte-Carlo simulation (cf.\ Sec.~\ref{sec:testing-3d-4d}),
and use this to estimate the expected sensitivity depth for each setup.
We use a canonical value of $\pfa=\num{e-10}$ (as was done in \cite{BinaryWeave:method}) for the
single-template false-alarm probability, which represents a somewhat typical false-alarm scale for wide
parameter-space searches. We quote the sensitivity depth for $\pdet=\SI{90}{\percent}, \SI{95}{\percent}$ and
$\SI{99}{\percent}$.
The former two are typical confidence-levels used for upper limits obtained in CW
searches, while the last one might be interesting, for example, if one is interested in rejecting the
torque-balance hypothesis or a specific emission mechanism in some parameter range at high confidence.

In Table~\ref{tab:search_setup}, we summarize the sensitivity depths for a set of six different search
setups. The sensitivity depths obtained from the empirical mismatch distributions corresponding to the
well-studied setup-I and setup-II are presented in this table. In addition, we report the maximum
achievable sensitivity depths for this \BinaryWeave{} pipeline estimated from our simulated searches for
four different setups that may be relevant for different cases of unknown spin wandering effect in Sco X-1.

\begin{table*}[htbp]
\begin{center}
\begin{tabular}{|l|c|c|c|c||c|c|c|}
 \hline
  Search setup & $T\obs$& $\Tseg$ & $\Nseg$ & $\mmax$ & $\depth_{\pfa}^{90\%}$ & $\depth_{\pfa}^{95\%}$ & $\depth_{\pfa}^{99\%}$ \\[1ex]
               & $[\si{months}]$ &$[\si{days}]$&     &         & $[\udepth]$ & $[\udepth]$ & $[\udepth]$ \\[1ex]
  \hline
  \setup{I}   & 6   &  1  & 180 & 0.031 & 77  &  72 &  60 \\
  \setup{II}  & 12  &  3  & 120 & 0.056 & 116 & 107 &  91 \\
  \hline
  \setup{III} & 6   &  3  &  60 & 0.025 & 96  &  89 &  75 \\
  \setup{IV}  & 12  &  1  & 360 & 0.025 & 93  &  86 &  73 \\
  \setup{V}   & 6   & 10  &  18 & 0.025 & 120 & 111 &  94 \\
  \setup{VI}  & 12  & 10  &  36 & 0.025 & 150 & 138 & 117 \\
 \hline
\end{tabular}
\end{center}
\caption{Definition of example search setups with corresponding estimated sensitivity depth, discussed in
  Sec.~\ref{sec:pred-sens-diff}.
  The sensitivity estimates assume a (per-template) false-alarm probability of $\pfa = \num{e-10}$ and
  detection confidences $\pdet = 90\%, 95\%,\text{and}\,99\%$, respectively, using the measured (4D)
  mismatch distributions obtained for each setup (cf.\ Sec.~\ref{sec:testing-3d-4d}).
}
\label{tab:search_setup}
\end{table*}

\subsection{Computing cost for different search scenarios}
\label{cost_fixed_depth}

\begin{table}[htbp]
  \begin{center}
    \input{table_search_cost-Mh.tex}
  \end{center}
  \caption{Computing-cost estimates $\Cost_{\Dop}$ (in million core hours [Mh]) for different parameter spaces
    $\Dop_n$ defined in Table.~\ref{tab:search_param_spaces}.
    We consider two setups, \setup{I} and \setup{II} of Table~\ref{tab:search_setup}, assuming either a 3D or
    4D template-bank.
  }
  \label{tab:search_costs}
\end{table}
Here we present CPU computing cost in terms of \emph{core hours}, and million
core hours (Mh), referring to the mix of CPU hardware used in the present study, cf.\ Sec.~\ref{cpu_runtime}.
Another interesting unit used in \cite{BinaryWeave:method} is Einstein@Home months (EM), which was defined as
\num{12000} (average) CPU cores running on Einstein@Home \cite{EinsteinAtHome} for 30 days.
If one assumes the (current) average Einstein@Home CPU to be roughly comparable to the one used here, one can
convert $\SI{1}{EM} \approx \SI{8.6}{Mh}$.
\useMSU{A measure of computing cost used for clusters in the LIGO Scientific Collaboration is the so-called
\emph{service unit} (SU), which refers to one core hour specifically on an {\tt Intel Xeon E5-2670} processor.
One can obtain a detailed SU conversion factors for different CPUs, and for the hardware mix used in this
study we find a conversion factor of $\SI{1.46}{SU/h}$.}

Let us first consider the example of the \Sco{} parameter space $\Dop_1$ considered in
\citet{BinaryWeave:method} (cf.\ Table~\ref{tab:search_param_spaces}) with two different search setups (I and
II) of Table.~\ref{tab:search_setup}.
For \setup{I} with $180\times\SI{1}{\day}$ segments and mismatch $\mmax=0.031$, the total number of (4D)
templates given by Eq.~\eqref{eqn:number_templates_4D} is $\N_{4\dims} = \num{5.84e14}$.
Using the effective computing-cost model of Eq.~\eqref{eq:8} this results in a total CPU runtime of
$\Cost_{\Dop_1}[\setup{I}] \approx \SI{8.46e10}{\second} = \SI{23.5}{Mh}$. Using the above conversion factors, this would
correspond to $\SI{2.7}{EM}$\useMSU{ or $\SI{34.3}{MSU}$}.
Similarly, for \setup{II} with $120\times\SI{3}{\day}$ segments and mismatch of $\mmax=0.056$,
we obtain a template count of $\N_{4\dims} = \num{1.07e15}$ and a corresponding total CPU runtime of
$\Cost_{\Dop_1}[\setup{II}]\approx \SI{1.56e11}{\second} = \SI{43.2}{Mh}$, which we can also express as
$\SI{5.0}{EM}$\useMSU{ or $\SI{63}{MSU}$}.

Next we consider a number of additional parameter-space scenarios, listed in Table.~\ref{tab:search_param_spaces}.
The constraints from optical and radio emission observations come from different sources in the literature
\cite{ScoX1OrbPar:BradshawEtal1999, ScoX1OrbPar:FomalontEtal2001}, with the most recent values given
in~\cite{Galloway:2014ApJ, ScoX1OrbPar:WangEtal2018}.
Future observations are likely to further alter and improve these constraints.
For a fully-resolved period uncertainty, the total number of templates (and therefore computing cost) scales
as $\asini{}_{\max}^{3}$ for a wide parameter uncertainty in $\asini$ (cf.\
Eq.~\eqref{eqn:number_templates_4D}), but only as $\asini{}_{\max}^{2}\,\Delta\asini$ for narrow parameter uncertainty $\Delta\asini$.

In order to quantify the effects of future improved constraints on $\asini$, we consider three different
scenarios:
(i) $\asini \in [1.45, 3.25]\,\si{ls}$ (search spaces $\Dop_4-\Dop_9$),
(ii) $\asini \in [1.40, 1.50]\,\si{ls}$ (search spaces $\Dop_{10}-\Dop_{15}$) and
(iii) $\asini \in [1.44, 1.45]\,\si{ls}$ (search spaces $\Dop_{16}-\Dop_{21}$).
Similarly we consider six different frequency search ranges, three ``deep-search'' ranges covering only
$\SI{100}{Hz}$ at different frequencies ($\SIrange{600}{700}{Hz}$, $\SIrange{1000}{1100}{Hz}$, and
$\SIrange{1400}{1500}{Hz}$), and three ``broad-search'' ranges within the LIGO/Virgo frequency band
($\SIrange{20}{500}{Hz}$, $\SIrange{20}{1000}{Hz}$ and $\SIrange{20}{1500}{Hz}$).
Finally, we consider both a 3D (for an unresolved period uncertainty $\Delta\Porb$) and 4D search for all
cases considered.

The resulting computing cost estimates for all combinations of the two setups (I and II), 3D or 4D template
bank, and different parameter spaces $\Dop_{1-21}$ are given in Table.~\ref{tab:search_costs}.
We note that while some required computing budgets may seem unrealistically large, a recent GPU port
of the $\F$-statistic and \Weave{}~\cite{DunnEtAl2022:GPrUImpFsCnGrvWS,Wette_etal_2021PRD} may yield speedups
factors of tens to hundreds, making many more setups fall within reach of currently available computing resources.

\subsection{Sensitivity versus computing cost}
\label{depth_fixed_cost}

\begin{figure*}[htbp]
  \includegraphics[width=\columnwidth]{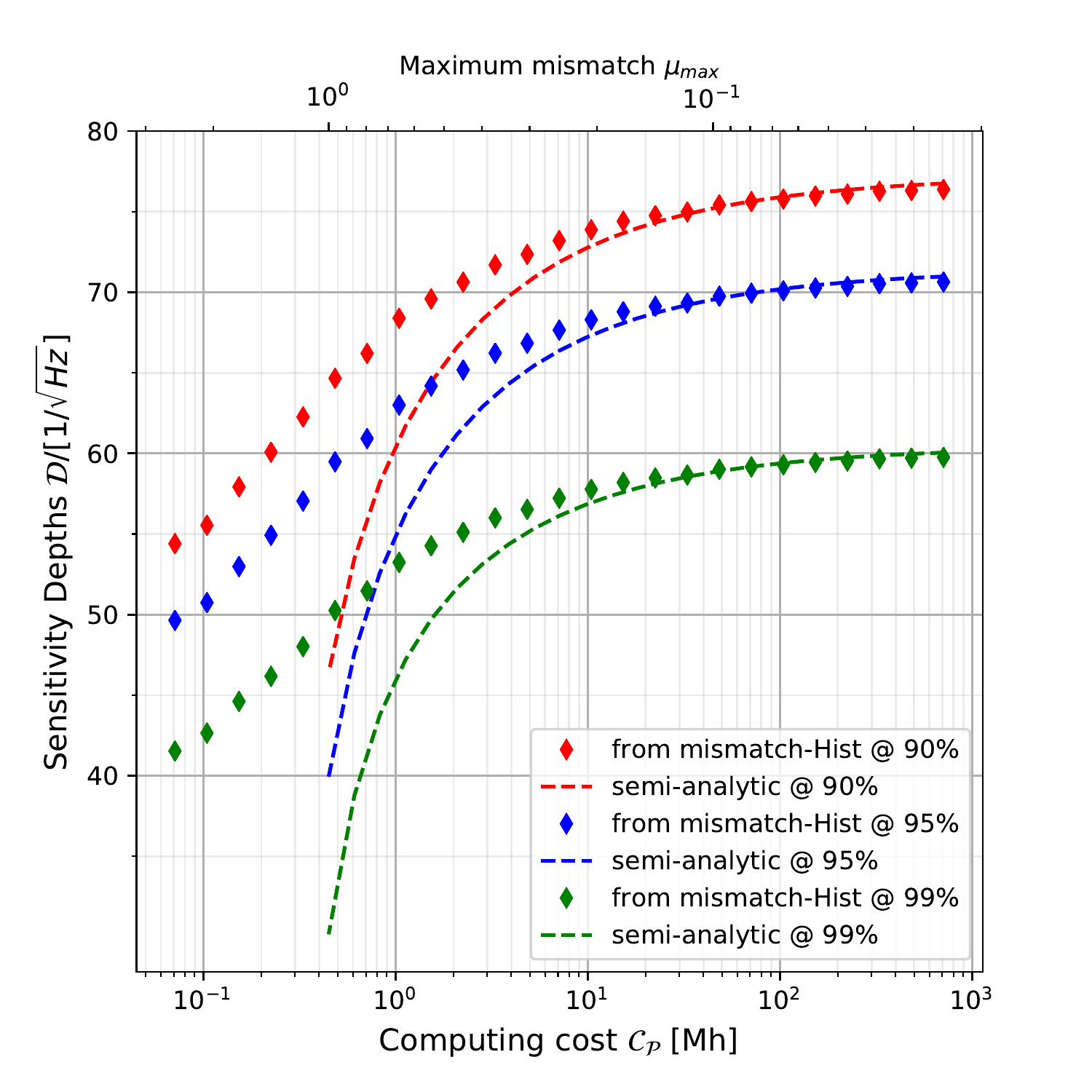}
  \includegraphics[width=\columnwidth]{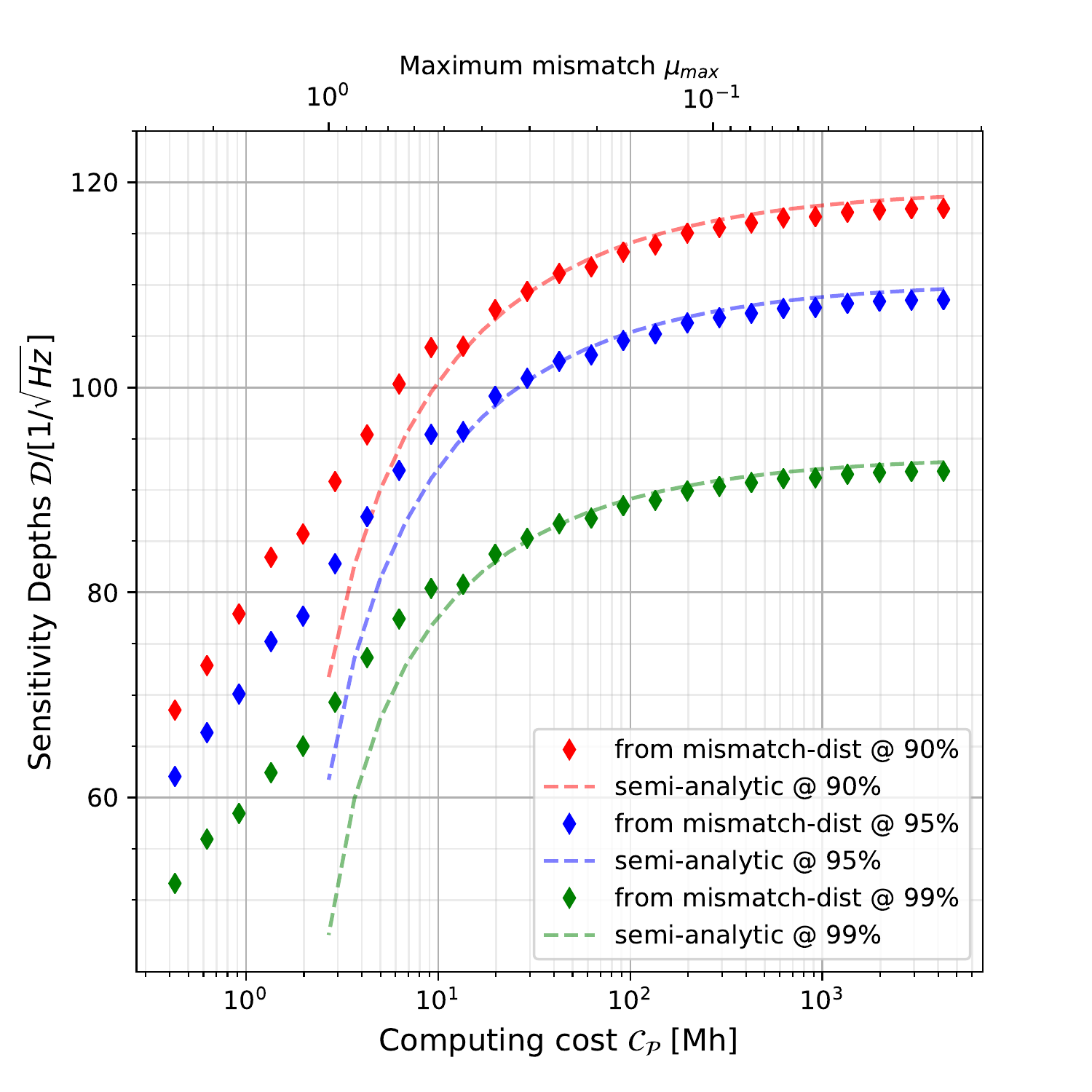}\\
  \caption{Sensitivity depth $\depth_{\pfa}^{\pdet}$ as a function of (4D) computing cost $\Cost_{\Dop}$ for
    varying maximum mismatch $\mmax$ at fixed segments ($\Nseg,\Tseg$), assuming \Sco{} parameter space $\Dop_2$ of
    Table~\ref{tab:search_param_spaces}.
    Sensitivity depth is estimated for a fixed (per-template) false-alarm of $\pfa = \num{e-10}$ and different
    confidence levels of $\pdet=\SI{90}{\percent}$ (top), $\pdet=\SI{95}{\percent}$ (middle) and $\pdet=\SI{99}{\percent}$ (bottom).
    Segment setup is $\Nseg\times\Tseg = 180\times\SI{1}{\day}$ (left plot), and $120\times\SI{3}{\day}$ (right plot),
    corresponding to \setup{I} and \setup{II}, respectively (cf.\ Table~\ref{tab:search_setup}).
    The dashed lines correspond to the sensitivity estimate assuming a theoretical \Ans{} lattice
    mismatch-distribution, while the diamond markers correspond to using the measured \BinaryWeave{} mismatch distributions.
    Computing cost is measured in million core hours (Mh).
  }
  \label{fig:sensdepth_computing_cost}
\end{figure*}
In addition to considering various fixed search scenarios as in the previous two subsections, it is also
instructive to study how the achievable sensitivity varies as a function of the invested computing cost.
This would generally involve a (3- or 4-dimensional) optimization problem over all search-setup parameters
(see \cite{PrixShaltev2011:optimalStackSlide,BinaryWeave:method}) which is beyond the scope of this study, so
we consider a simpler problem of varying the maximal template-bank mismatch $\mmax$.
In a sense, this provides a \emph{lower limit} on the achievable sensitivity at any given cost, as one could
always improve sensitivity further by varying all three setup parameters $\{\mmax,\Nseg,\Tseg\}$ at fixed cost.

The search space is chosen as $\Dop_{2}$, and we use again \setup{I} (i.e.,
$180\times\SI{1}{\day}$ segments) and \setup{II} (i.e., $180\times\SI{3}{\day}$) as baselines, but now we vary
the maximal template-bank mismatch in the range $0.025 \leq \mmax \leq 2.5$.
For each mismatch, we can estimate the number of templates $\N_{4D}\propto \mmax^{-2}$ via
Eq.~\eqref{eqn:number_templates_4D}, and obtain the corresponding computing cost $\Cost$ from the simplified
cost model Eq.~\eqref{eq:8}.
We use the corresponding theoretical mismatch distribution\footnote{This will be a
  conservative over-estimate of the mismatch, see Sec.~\ref{sec:testing-3d-4d}, and therefore an
  under-estimate of the sensitivity.} for the \Ans{}-lattice, as well as the measured distribution from a set
of \num{100} injection-recovery simulations using \BinaryWeave{}, to estimate the expected sensitivity depth
via \texttt{SensitivityDepthStackSlide()} from \textsc{OctApps}.

This allows us to plot sensitivity depth versus computing cost, parametrized along $\mmax$ at fixed segment
setup $\Nseg\times\Tseg$, which is shown in Fig.~\ref{fig:sensdepth_computing_cost}.
As expected, sensitivity improves as the invested computational cost increases and (equivalently) the
maximum mismatch decreases; for $\mmax \lesssim 0.1$, however, further gains in sensitivity are minimal.
We observe good agreement at small mismatches (i.e., large computing costs) between the theoretical estimates
(using expected lattice mismatch distributions) and estimates using the measured mismatch distributions.
The small loss of the measured versus expected sensitivity in this regime from (well known) additional
intrinsic losses ($\sim\Ord{\SIrange{1}{3}{\percent}}$) of the high-performance $\F$-statistic implementation
compared to the exact calculation.
At higher mismatches $\mmax$, the measured mismatches tend to be smaller than the metric predictions, due to
neglected higher-order terms in the metric approximation, as discussed previously in
Sec.~\ref{subsec:template_mismatch} and Sec.~\ref{sec:testing-3d-4d}.
This explains the measured sensitivity decreasing more slowly compared to the theoretical estimates at higher
mismatches (i.e., smaller computing cost).

\section{Summary and outlook}
\label{sec:summary-outlook}

In this paper, we presented the implementation and characterization of \BinaryWeave{}, a new
semi-coherent search pipeline for CWs from neutron stars in binary systems with
known sky-position.
This pipeline is based on the \Weave{} framework \cite{IsolatedWeave}, initially developed for all-sky
searches of isolated sources, using the well established semi-coherent StackSlide $\Fsco$-statistic.

The \Weave{} framework requires a constant metric over the search parameter space for lattice tiling, and in
order to apply the non-constant binary metric of \citet{BinaryWeave:method}, we needed to develop a new
internal coordinate system in which a constant approximation to the binary metric can be obtained.
This is the basis for the \BinaryWeave{} implementation.
We performed extensive Monte-Carlo tests for the safety (in terms of mismatches) of the resulting template
banks and their template counts versus theoretical model expectations.
Furthermore, we obtained a simplified timing model for the non-interpolating StackSlide mode used here, which
allows easy estimates for the required computing cost of a given search, based on the known analytic
template-count models.

Putting these pieces together, we illustrate expected sensitivity depths for \BinaryWeave{} assuming different
search setups, and we estimate the corresponding required computing costs for a number of different \Sco{}
parameter-space regions of interest.

Two other primary pipelines, \texttt{CrossCorr} and \texttt{Viterbi}, are presently used for searching
CW-signals from Sco X-1. \texttt{Viterbi} pipeline aims to track the stochastic phase evolution model
due to spin-wandering effect of the neutron star in Sco X-1. It is thus more robust against this effect.
The computational cost is also quite less compared to the most sensitive searches of BinaryWeave. However,
the maximum achievable sensitivity depth for Viterbi is also less as compared to the most sensitive search
of BinaryWeave provided the spin-wandering effect is not significantly large. 

The sensitivity of \texttt{CrossCorr} pipeline is expected to be comparable to \BinaryWeave{}. The
computing cost for \emph{resampling} \texttt{CrossCorr} is also expected to be comparable to 
\BinaryWeave{}. However, \BinaryWeave{} can be adopted to utalize different grid spacing for coherent
and semi-coherent template banks that can reduce the computing cost for a search. This extra amount of
computing resource can be reutilized to further increase the sensitivity depth of \BinaryWeave{} by either
decreasing the mismatch or increasing the segment lengths.

One of the primary goals of developing \BinaryWeave{} is to perform searches for \Sco{} that can beat the
torque-balance limit over as wide a frequency range as possible, and are able to take advantage of any 
large available computing budget. Still, at the current level of electromagnetic constraints on the \Sco{} 
parameters, reaching the torque-balance limit over the full frequency range remains computationally 
prohibitive. Future improvements in these constraints will be immensely impactful to increase the chances
of detecting a CW signal from \Sco{} (or other LMXBs), as illustrated in Sec.~\ref{cost_fixed_depth}.

\begin{acknowledgments}

  AM acknowledges Stuart Anderson, James Clark, Duncan Macleod, Dan Moraru, Keith Riles, Peter Shawhan and
  several other members in computing and software team of the LIGO Scientific Collaboration (LSC). AM is
  thankful to Heinz-Bernd Eggenstein for learning some of the advanced computational skills. AM also
  acknowledges computational assistance by Henning Fehrmann and Carsten Aulbert. AM is thankful to Grant
  David Meadors and several other past and present members of the continuous-waves working group of the
  LSC regarding general discussion on detectibility of CW signal from Sco X-1. We thank Pep Covas and
  Paola Leaci for helpful feedback on the manuscript.

  This work has utilized the LDAS computing clusters at the LIGO Hanford Observator (LHO) CalTech LIGO
  centre (CIT) and the ATLAS computing cluster at the MPI for Gravitational Physics Hannover. AM acknowledges
  support from the DST-SERB Start-up Research Grant SRG/2020/001290 for completion of this project. KW was
  supported by the Australian Research Council Centre of Excellence for Gravitational Wave Discovery (OzGrav)
  through project number CE170100004.

\end{acknowledgments}

\bibliography{BinaryWeave}

\end{document}

%% file: defs.tex
\usepackage{siunitx}

\newcommand{\tref}{t_{\mathrm{ref}}}
\newcommand{\subp}{_\mathrm{p}}
\newcommand{\asini}{a\subp}
\newcommand{\kappap}{\kappa\subp}
\newcommand{\etap}{\eta\subp}
\newcommand{\tp}{t\subp}
\newcommand{\vp}{v\subp}
\newcommand{\orb}{_{\mathrm{orb}}}
\newcommand{\Porb}{P\orb}
\newcommand{\asc}{_{\mathrm{asc}}}
\newcommand{\sft}{_{\mathrm{sft}}}
\newcommand{\obs}{_\mathrm{obs}}
\newcommand{\pfa}{p_{\mathrm{fa}}}
\newcommand{\pdet}{p_{\mathrm{det}}}
\newcommand{\ma}{_{\mathrm{ma}}}
\newcommand{\mal}{_{\mathrm{ma},\ell}}
\newcommand{\F}{\mathcal{F}}
\newcommand{\rot}{_{\mathrm{rot}}}
\newcommand{\Sco}{\mbox{Sco X-1}}
\newcommand{\A}{\mathcal{A}}
\newcommand{\coh}[1]{{#1}}
\newcommand{\sco}[1]{\hat{#1}}
\newcommand{\Fcoh}{\coh{\F}}
\newcommand{\Fsco}{\sco{\F}}
\newcommand{\sig}{_{\mathrm{s}}}
\newcommand{\tmpl}{_{\mathrm{t}}}
\newcommand{\Dop}{\mathcal{P}}
\newcommand{\Nseg}{N}
\newcommand{\Tseg}{\Delta T}
\newcommand{\midl}{_{\mathrm{mid},\ell}}
\renewcommand{\mid}{_{\mathrm{mid}}}
\newcommand{\av}[1]{\overline{#1}}
\newcommand{\avg}[1]{\left\langle{#1}\right\rangle}
\newcommand{\Weave}{\textsc{Weave}}
\newcommand{\BinaryWeave}{\textsc{BinaryWeave}}
\newcommand{\mis}{\mu}
\newcommand{\mmax}{\mis_{\max{}}}
\newcommand{\mmaxcoh}{\coh{\mis}_{\max{}}}
\newcommand{\mmaxsco}{\sco{\mis}_{\max{}}}
\newcommand{\Ans}{$A_n^*$}
\newcommand{\Ord}[1]{\mathcal{O}\left(#1\right)}
\newcommand{\fix}[1]{\underline{#1}}
\newcommand{\dims}{\mathrm{D}}
\newcommand{\N}{\mathcal{N}}
\newcommand{\setup}[1]{\texttt{search setup-#1}}
\newcommand{\eff}{^{\mathrm{eff}}}
\newcommand{\Cost}{\mathcal{C}}
\newcommand{\depth}{\mathcal{D}}
\newcommand{\Sn}{S_{\mathrm{n}}}
\newcommand{\udepth}{\si{1\per\sqrt\hertz}}

%% file: table_search_cost-Mh.tex
\begin{tabular}{lrrrr}
\toprule
{} &  (I,3D) &   (I,4D) &  (II,3D) &  (II,4D) \\
\midrule
$\Dop_{1}$  &    3.18 &    23.51 &     3.93 &    43.23 \\
$\Dop_{2}$  &   28.50 &   466.48 &    35.22 &   857.69 \\
$\Dop_{3}$  &    5.00 &    63.40 &     6.17 &   116.57 \\
$\Dop_{4}$  &   26.38 &   577.57 &    32.60 &  1061.95 \\
$\Dop_{5}$  &   68.76 &  2425.79 &    84.96 &  4460.17 \\
$\Dop_{6}$  &  131.09 &  6381.48 &   161.97 & 11733.30 \\
$\Dop_{7}$  &    3.24 &    20.42 &     4.01 &    37.54 \\
$\Dop_{8}$  &  207.74 &  5226.87 &   256.69 &  9610.37 \\
$\Dop_{9}$  &  701.14 & 26461.02 &   866.33 & 48652.49 \\
$\Dop_{10}$ &    0.90 &    11.65 &     1.12 &    21.42 \\
$\Dop_{11}$ &    2.36 &    48.94 &     2.91 &    89.97 \\
$\Dop_{12}$ &    4.49 &   128.73 &     5.55 &   236.70 \\
$\Dop_{13}$ &    0.11 &     0.41 &     0.14 &     0.76 \\
$\Dop_{14}$ &    7.12 &   105.44 &     8.80 &   193.87 \\
$\Dop_{15}$ &   24.03 &   533.80 &    29.70 &   981.46 \\
$\Dop_{16}$ &    0.09 &     1.16 &     0.11 &     2.13 \\
$\Dop_{17}$ &    0.23 &     4.86 &     0.29 &     8.93 \\
$\Dop_{18}$ &    0.45 &    12.78 &     0.55 &    23.50 \\
$\Dop_{19}$ &    0.01 &     0.04 &     0.01 &     0.08 \\
$\Dop_{20}$ &    0.71 &    10.47 &     0.88 &    19.25 \\
$\Dop_{21}$ &    2.40 &    52.99 &     2.96 &    97.43 \\
\bottomrule
\end{tabular}